\documentclass[final]{elsarticle}
\makeatletter
\def\ps@pprintTitle{%
 \let\@oddhead\@empty
 \let\@evenhead\@empty
 \def\@oddfoot{\centerline{\thepage}}%
 \let\@evenfoot\@oddfoot}
\makeatother
\usepackage{lineno,hyperref}
\usepackage{amsmath} 
\usepackage{amsfonts}
\usepackage{mathtools}

\nolinenumbers
\usepackage{makeidx}
\usepackage{multirow}
\usepackage{multicol}
\usepackage[dvipsnames,svgnames,table]{xcolor}
\usepackage{graphicx}
\usepackage{epstopdf}
\usepackage{ulem}
\usepackage{amssymb}
\DeclareMathAlphabet{\mathbit}{OT1}{cmr}{bx}{it}

\newtheorem{thm}{\underline{Theorem}}
\newtheorem{prop}{\underline{Proposition}}

\newproof{pf}{Proof}

\usepackage{subcaption}
\usepackage{color, colortbl}
\definecolor{Gray}{gray}{0.9}
\definecolor{Vio}{rgb}{0.5,0,1}
\definecolor{Org}{rgb}{1,0.2,1}
\definecolor{LightCyan}{rgb}{0.88,1,1}
\definecolor{LightRed}{rgb}{1,0.88,0.88}
\definecolor{LightBlue}{rgb}{0.12,0.7,0.8}
\definecolor{LighterBlue}{rgb}{0.12,0.9,0.9}
\definecolor{Red}{rgb}{1,0,0}
\definecolor{Green}{rgb}{0,0.5,0}
\definecolor{LightGreen}{rgb}{0,1,0}
\usepackage[table]{xcolor}
\usepackage{tabulary}
\usepackage{tabularx}

\usepackage{booktabs} 
\usepackage{xparse}   
\NewDocumentCommand{\rot}{O{45} O{1em} m}{\makebox[#2][l]{\rotatebox{#1}{#3}}}%

\begin{document}

\begin{frontmatter}

\title{To schedule or not to schedule: \\when no-scheduling can beat the best-known flow scheduling algorithm in datacenter	 networks}

\author{Soheil Abbasloo$^1$}
\author{Yang Xu$^2$}
\author{H. Jonathan Chao$^1$}
\address{ab.soheil@nyu.edu, xuy@fudan.edu.cn, chao@nyu.edu}
\address{New York University$^1$, New York, USA}
\address{Fudan University$^2$, Shanghai Shi, China}
%

\begin{abstract}
Conventional wisdom for minimizing the average flow completion time (AFCT) in the datacenter network (DCN), where flow sizes are highly variable, would suggest scheduling every individual flow. However, we show that considering scheduling delay (including scheduler’s computational and communication delays), serving most of the flows without any scheduling and only in first-come-first-served (FCFS) manner significantly improves their performance even when it is compared to the shortest remaining processing time (SRPT)–known as optimum algorithm when scheduling delay is zero. To do so, we only require to have two coarse classes of flows categorized based on flows’ sizes (1st-class including flows smaller than a threshold, H, and 2nd-class including others) and serve 1st-class flows always before serving 2nd-class ones.
To show that, we take SRPT scheduling algorithm accompanied by the global knowledge of flows, formulate impact of scheduling delay on its performance, and prove that for any flow size distribution and network load ($<1$), there is always a threshold, H, which guarantees 1st-class flows achieve lower AFCT under FCFS compared to SRPT. Our numerically calculated results and extensive flow-level simulations show that on average, more than 90\% of flows could be in 1st-class and consequently do not require any scheduling.

\end{abstract}

\begin{keyword}
flow scheduling \sep datacenter network \sep flow completion time \sep SRPT \sep FCFS
\end{keyword}

\end{frontmatter}


\section{Introduction}
In the network context, there is a vast amount of work relevant to scheduling from the simplest solutions such as first-come-first-served (FCFS) or last-come-first-served (LCFS) to more complicated algorithms such as shortest remaining processing time (SRPT). Considering the average completion time as the objective, since long time ago, it has been well-known that SRPT is the optimum scheduling solution over a single-link~\cite{opt} which always prioritizes flows with smaller remaining sizes over others and always serves the highest priority flows first in the network.

The response time of today’s popular datacenter applications such as web search, social networks, and recommendation systems highly impacts the end-user satisfaction and consequently total revenue of these interactive applications~\cite{rev}. This motivates huge body of work recently proposing new datacenter scheduling designs to minimize average flow completion time (AFCT) as the primary objective determined mainly by the end-to-end latency of datacenter networks (DCNs) (e.g.~\cite{pdq,fastpass,pfabric,pase}). Due to the optimality of SRPT algorithm over single-link, most of these schemes use SRPT (or its approximations) in the core of their scheduling designs to schedule all flows in the network so that they can achieve lower AFCT (e.g.~\cite{pdq,fastpass,pfabric,pase}). To that end, they gather a global view of the network’s flows’ information (such as their sizes and source-destination addresses) through either centralized (e.g. ~\cite{fastpass,pase}) or distributed (e.g.~\cite{pdq}) approaches, and schedule flows based on their global priorities. Also, today’s growing interest and advances in using the centralized approaches for monitoring and controlling the network (such as SDN) help these schemes to become closer to their goals.

However, in practice, scheduling flows using SRPT by having a global view of the network’s flows and their information, comes always with a cost. More precisely, getting a global view of the network’s flows at least requires a round trip time (RTT) delay. This delay in the centralized approach could be from end-hosts (or the switches) to the logically centralized scheduler. In distributed schemes, this delay depending on the main strategy could be the delay for negotiations between end-hosts and switches, among switches, or end-hosts. Add to it the computational delay of the scheduler. So, in DCNs where we have lots of end-hosts communicating with each other, we will always experience scheduling delay. In contrast with a general network, this delay becomes important in the DCN context, where a lot of small flows can finish ideally just in a few RTTs~\cite{dctcp,vl2}.

Therefore, in this paper, instead of adding another scheme to the impressive collection of scheduling proposals for DCNs, we challenge the conventional wisdom suggesting that to minimize AFCT, every individual flow should be scheduled (fine-grained scheduling). To that end, we model the best-known scheduling algorithm, SRPT, accompanied by the global view of the flows’ information when scheduling delay (delay of getting information of all network’s flows and computational delay at scheduler) is not zero. Having that model, we analyze and formulate the impact of scheduling delay on the overall performance of the system. Based on those analysis, we show that in DCNs, when flows are categorized into two coarse classes based on their sizes (1st-class including flows smaller than a threshold and 2nd-class including others) and when all 1st-class flows receive service before 2nd-class ones (simply by having two queues and using strict priority scheme to serve these queues), serving 1st-class flows in FCFS manner (i.e., using no-scheduling) leads to lower AFCT for 1st-class flows compared to using SRPT algorithm for them.
Interestingly, we show that considering various workloads (including realistic production datacenter workloads and synthetic workloads) and different traffic loads ($<1$), on average, more than 90\% of flows could be placed in the 1st-class category in which, flows are not scheduled individually and simply receive service in FCFS manner. 

We believe that our analysis and results will provide a solid ground for the design of new coarse-grained scheduling schemes in DCNs which are rooted in the fact that fine-grained scheduling approaches with a global knowledge of network’s flows are always suboptimal.

In this paper, we make three main contributions:
\begin{itemize}
\item We analyze and formulate the impact of scheduling delays on the overall performance of the scheduler.
\item We show that to minimize the AFCT in DCNs, in contrast with the conventional wisdom suggesting the necessity of doing fine-grained scheduling (i.e., scheduling every individual flow), most of the flows should not be scheduled.
\item We prove that for any flow size distribution and traffic load ($<1$), for classifying flows, there is always a threshold, H, that guarantees 1st-class flows achieve lower AFCT under FCFS compared to SRPT.
\end{itemize}

The rest of this paper is organized as follows. After discussing the main motivations in section II, section III provides the model and analysis to address the question of whether fine-grained scheduling should be done in DCNs or simply serving most of the flows in FCFS manner suffices. Section IV is dedicated for the evaluation including numerically calculated results and output results of extensive flow-level simulations. Finally, after stating the related work, we conclude the paper.

\section{Motivation}
Since a long time ago it has been known that SRPT algorithm has the lowest mean completion time of any scheduling algorithm under any flow arrival and flow size distributions over a single link~\cite{opt}. However, in production DCNs there are at least two key issues discussed as follows.

\subsection{Cost of Scheduling and Global-Awareness ($T_{cost}$)}
 One of the main properties of SRPT is that it needs to know the information of all flows coming to the network. In other words, the remaining size of all flows should be known during the scheduling~\cite{mil}. Although this might not be a major concern when there is only one single link in the network, in a network such as DCN which connects hundreds or thousands of machines this becomes an issue. Recently growing interest and advances in using centralized structures and techniques (such as a software-defined network (SDN)) to control/monitor the entire network enables designers to have access to the information of whole network’s flows in a logically centralized entity. That’s one of the reasons why there is a wide range of scheduling proposals in DCNs using SRPT (or its approximations) as the core of their centralized designs (e.g.~\cite{pdq,fastpass,pase}). However, getting a global view of the flows and their information requires time, and this delay could impact the overall performance of the scheduler. Note that this concern is not only for centralized schemes but also for the distributed ones. For instance,~\cite{pdq} which takes a distributed approach and uses SRPT in its core, passes flows’ information and their requests to different switches on the path to their destinations, and later uses the overall decision made by switches to take proper actions. Therefore, despite taking centralized or distributed approaches for scheduling flows, global-awareness always comes with a cost. Also, in a network such as DCN with hundreds of thousands of flows to be scheduled, the computational delay could be another issue. 
 
\subsection{DCNs’ Special Characteristics}

Real DCNs have important characteristics that distinguish them from general networks. One of these main characteristics is their traffic patterns. Traffic analysis of today’s DCNs illustrates that DCNs’ workloads include a wide range of flow sizes (e.g.~\cite{dctcp,vl2}). For instance,~\cite{vl2} shows that data mining applications have flow sizes from less than 1KB to bigger than 950 MB. As another example, studies in~\cite{dctcp} show that in a web search workload, over 95\% of all bytes are from only the largest 30\% of flows with sizes in the range of 1MB to 30 MB. Based on these studies, a common characteristic in DCNs’ workloads is that most flows are small while they account for a small portion of all bytes transferred in DCNs. Another important property of DCNs is that the round trip time delay compared to a general network is very low ($\approx100\mu s$). So, considering traffic characteristic and low RTT in DCNs, we could expect that most flows could ideally finish in a few RTTs.

Putting all together, any scheduling delay even as low as one RTT could impact the overall performance of scheduling schemes in DCNs. This motivates us to stop and instead of proposing just another scheduling design in DCN, go back one step and challenge the conventional wisdom that suggests scheduling every individual flow to minimize AFCT.

\section{To Schedule or Not To Schedule}
\subsection{Background on Mean-Analysis of M/G/1/SRPT and M/G/1/FCFS Queues}
Here, a brief background on the mean-analysis of the M/G/1/SRPT queue model based on expressions derived by~\cite{mil} is presented. In addition, we present mean response time of the M/G/1/FCFS queue model derived by Pollaczek-Kinchin~\cite{pk}. Considering a single-queue model of network, we denote the average arrival rate of the flows by $\lambda$. We assume that the flow size distribution is c.f.m.f.v. with probability density function $f(t)$. The cumulative flow size distribution is denoted by $F(t)$. $X$ refers to the service time of a flow. The total load is $\rho=\lambda\int_{0}^{\inf} tf(t) dt$, and the load made up by the flows with sizes less than or equal to $x$ is $\rho(x)=\lambda\int_{0}^{x} tf(t) dt$. We define $m_2(x)$ as follows: $m_2(x)=\int_{0}^{x} t^2f(t) dt$. The expected completion time for a flow of size $x$ using SRPT algorithm, $E[T(x)]_{SRPT}$, can be decomposed into the expected waiting time of the flow, $E[W(x)]_{SRPT}$, and the expected residence time of the flow $E[R(x)]_{SRPT}$. The waiting time of a flow is defined as the time from when it first arrives to when it receives service for the first time, and the residence time of a flow is the time from when it receives service for the first time to when its service is completed. The formulas for these expressions derived by~\cite{mil} are as follows:
\begin{align}
&E[T(x)]_{SRPT}=E[W(x)]_{SRPT}+E[R(x)]_{SRPT} \label{eq_1a} \\
&E[W(x)]_{SRPT}=\frac{\lambda(m_2 (x)+x^2 (1-F(x)))}{2(1-\rho(x))^2} \label{eq_1b} \\
&E[R(x)]_{SRPT}=\int_{0}^{x}\frac{dt}{(1-\rho(t))} 
\label{eq_1c}
\end{align}

So, the total mean completion time, $E\left[T\right]_{SRPT}$, is given by:
\begin{align}
&E\left[T\right]_{SRPT}=\int_{0}^{\infty}{E\left[T\left(x\right)\right]_{SRPT}}f\left(x\right)dx  \label{eq_1d}
\end{align}

For an M/G/1/FCFS queue, based on the Pollaczek–Khinchine formula~\cite{pk}, total expected completion time $E\left[T\right]_{FCFS}$, total expected waiting time, $E\left[W\right]_{FCFS}$, and total expected residence time, $E\left[R\right]_{FCFS}$, are given as:

\begin{align}
&E\left[T\right]_{FCFS}=E\left[W\right]_{FCFS}+E\left[R\right]_{FCFS}\label{eq_2a} \\
&E{\left[W\right]}_{FCFS}=\frac{\lambda{}E\left[X^2\right]}{2\left(1-\rho{}\right)}=\frac{\lambda{}\int_0^{\infty{}}t^2f\left(t\right)dt}{2\left(1-\rho{}\right)}\label{eq_2b} \\
&E{\left[R\right]}_{FCFS}=E\left[X\right]=\int_0^{\infty{}}xf\left(x\right)dx \label{eq_2c} 
\end{align}

We use Eq.~\ref{eq_1a},~\ref{eq_1b},~\ref{eq_1c},~\ref{eq_1d},~\ref{eq_2a},~\ref{eq_2b}, and~\ref{eq_2c} as the base for our next analysis in the rest of this paper.

\subsection{Modeling the Schedulers’ Structure in DCNs}

\begin{figure}[!t]
\center 
\includegraphics[width=0.63\linewidth,height=1.4in]{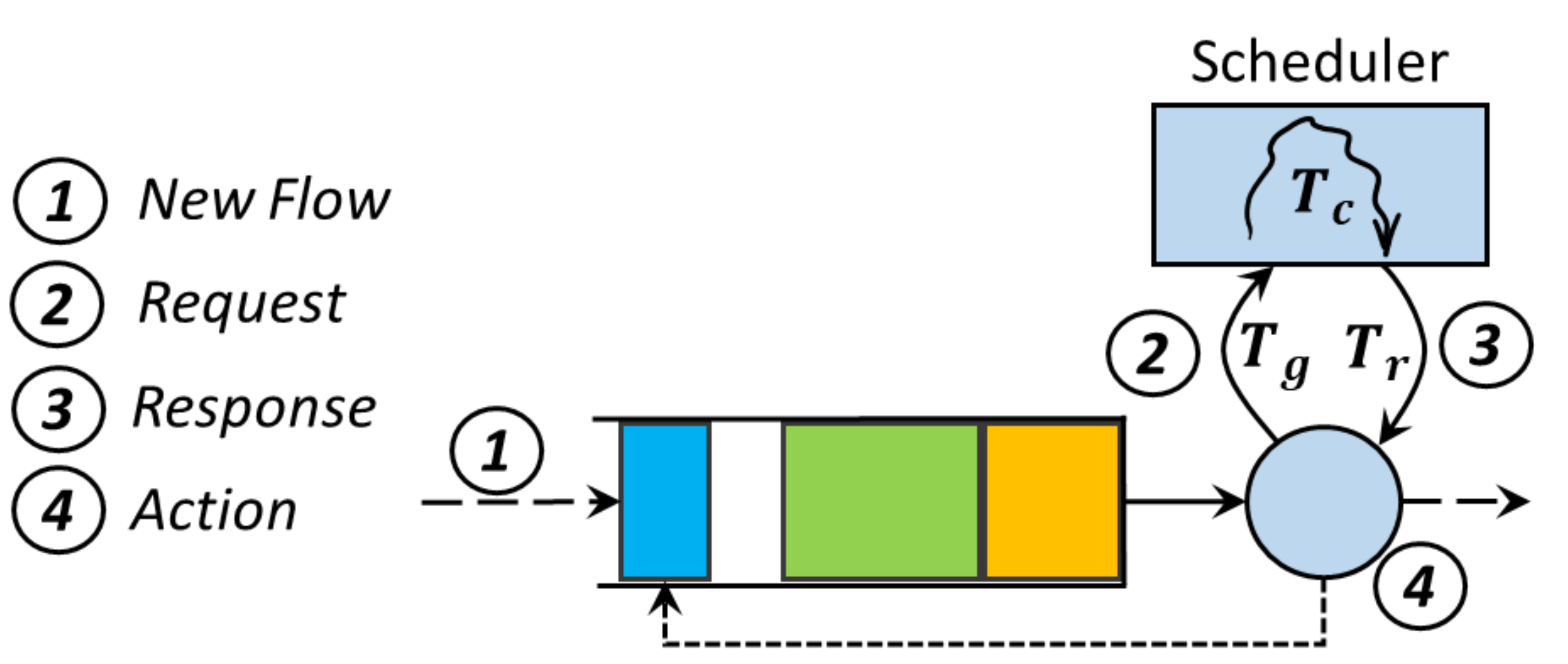}
\caption{Practical SRPT model considering three main delays: $T_g$, $T_c$, and $T_g$}    
\label{fig_srpt}
\end{figure}

Having the global knowledge of the network’s flows during the flow scheduling in DCN always requires negotiation between at least two entities in the network. In a distributed approach, end-hosts and switches might negotiate with each other or among themselves to find the highest priority flows (with shortest remaining sizes) and then serve them (e.g.~\cite{pdq}). In a centralized strategy, end-hosts (or switches) need to identify their flows and their corresponding remaining sizes to a central entity, receive the response and act based on that (e.g.~\cite{fastpass,pase}). Therefore, despite the choice of using a centralized or a distributed SRPT scheduler, getting a global knowledge of the network’s flows always introduces extra delays. We model the scheduler with scheduling delays considered in Fig.~\ref{fig_srpt}. Here, to simplify our analysis, we model the network as a single M/G/1 queue. 

Considering this model, there are at least 3 major delays seen in a scheduler in practice: 
\begin{itemize}
\item     Time delay for gathering flows’ information ($T_g$)
\item     Computational delay ($T_c$)
\item     Time delay for getting response either from a central entity or from some distrusted ones ($T_r$)
\end{itemize}

Since the scheduler is aware of the service rate (network’s links’ speeds) and when and how much each flow has been served, it is aware of the remaining sizes of all current flows (i.e., their priorities). Therefore, we do not need to consider the cost of updating scheduler’s knowledge about the current flows. With the same reasoning, the scheduler can calculate the finishing time of the flow receiving service at the current time ($f_1$). So, if for example, the scheduler should give service to the next highest priority flow ($f_2$) after serving $f_1$, it can send its response to the sender of $f_2$, $T_r$ timeslots sooner than finishing time of $f_1$. In practice, using similar approaches could avoid the extra delays of updating scheduler’s knowledge when there is no new flow coming to the network. That’s why here, delay for updating the knowledge of the scheduler is considered only when a new flow comes into the network.

\begin{figure}[!t]
\center 
\begin{minipage}[b]{0.49\linewidth}
\includegraphics[width=\linewidth,height=1.4in]{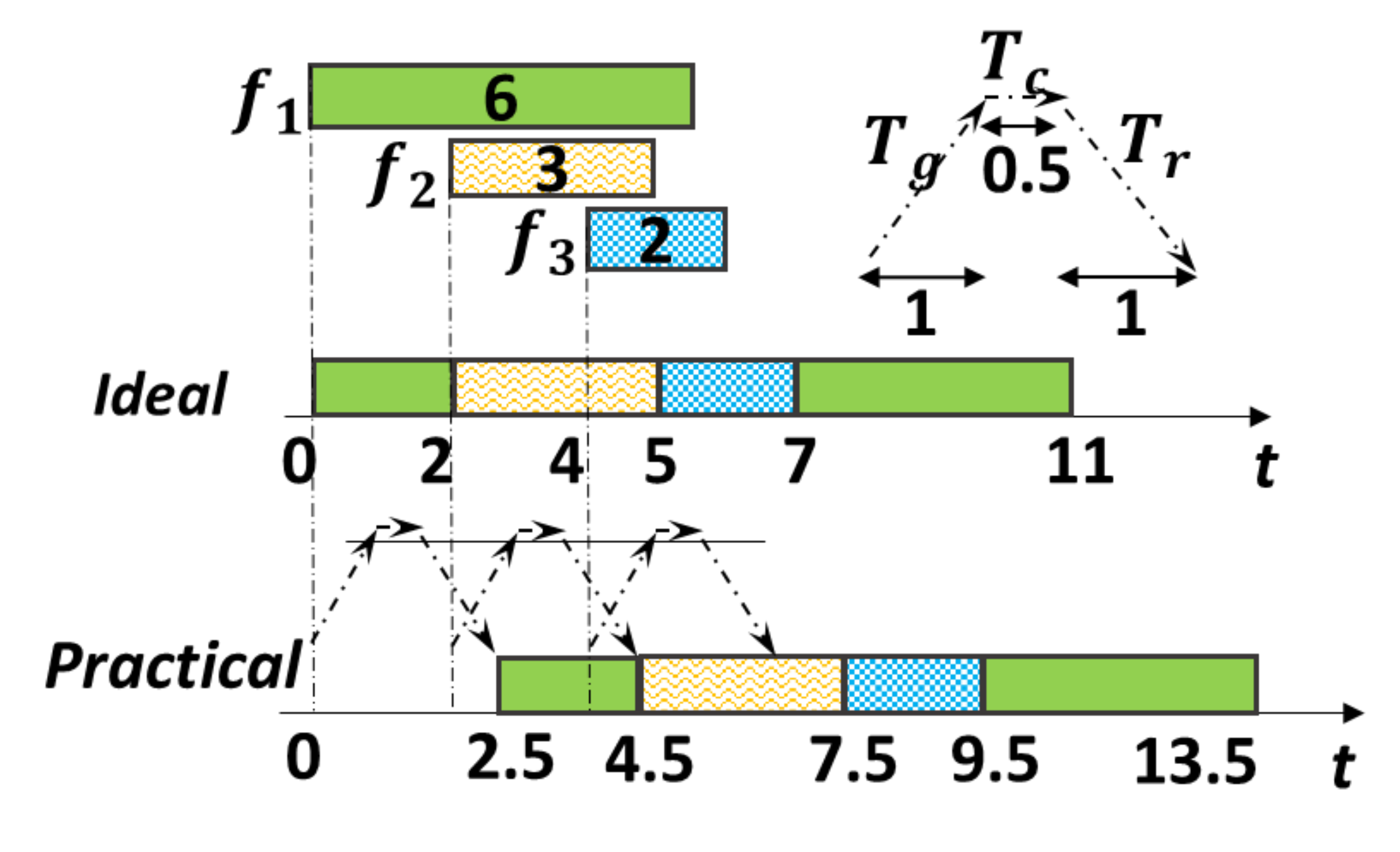}
\subcaption{Scheduler’s output in ideal and practical scenarioes}    
\end{minipage}
\hfill
\begin{minipage}[b]{0.49\linewidth}
    \begin{minipage}[b]{0.48\linewidth}
    \includegraphics[width=\linewidth,height=0.6in]{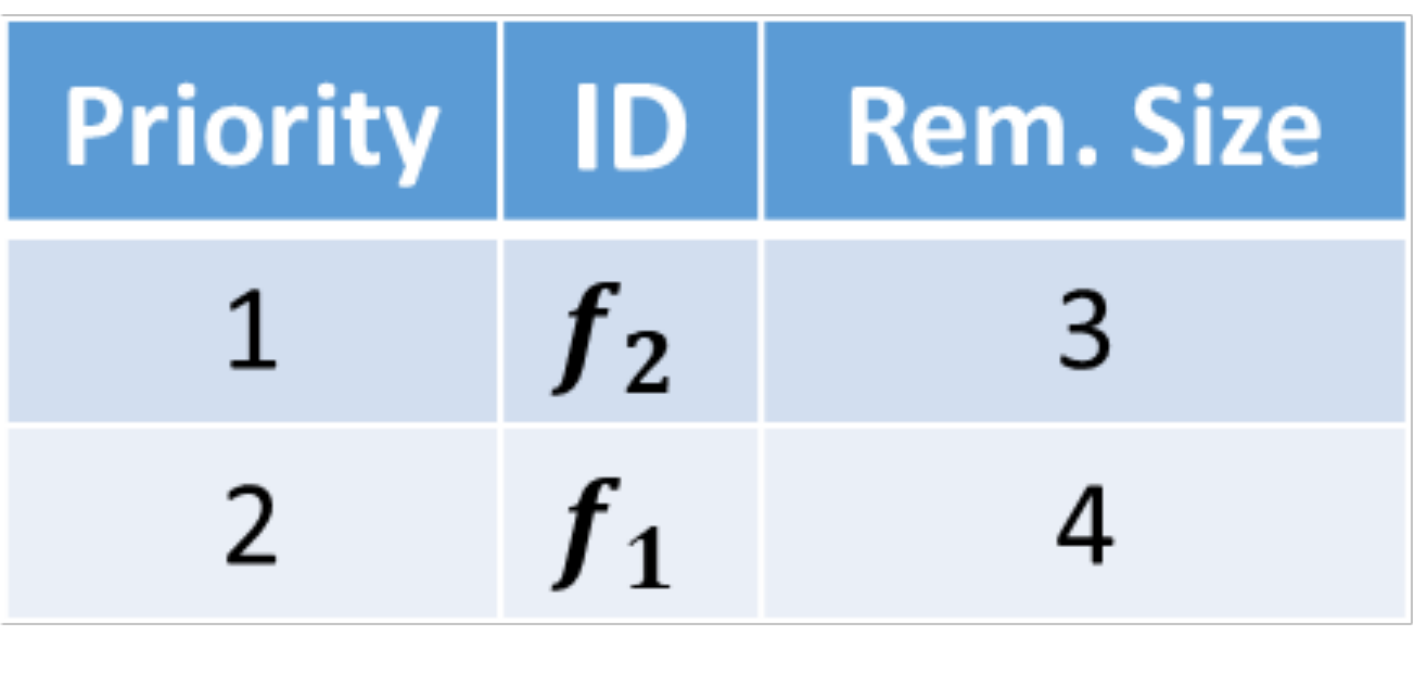}
    \subcaption{Scheduler’s knowledge at time 2 in ideal case}    
    \end{minipage}
    \hfill
    \begin{minipage}[b]{0.48\linewidth}
    \includegraphics[width=\linewidth,height=0.6in]{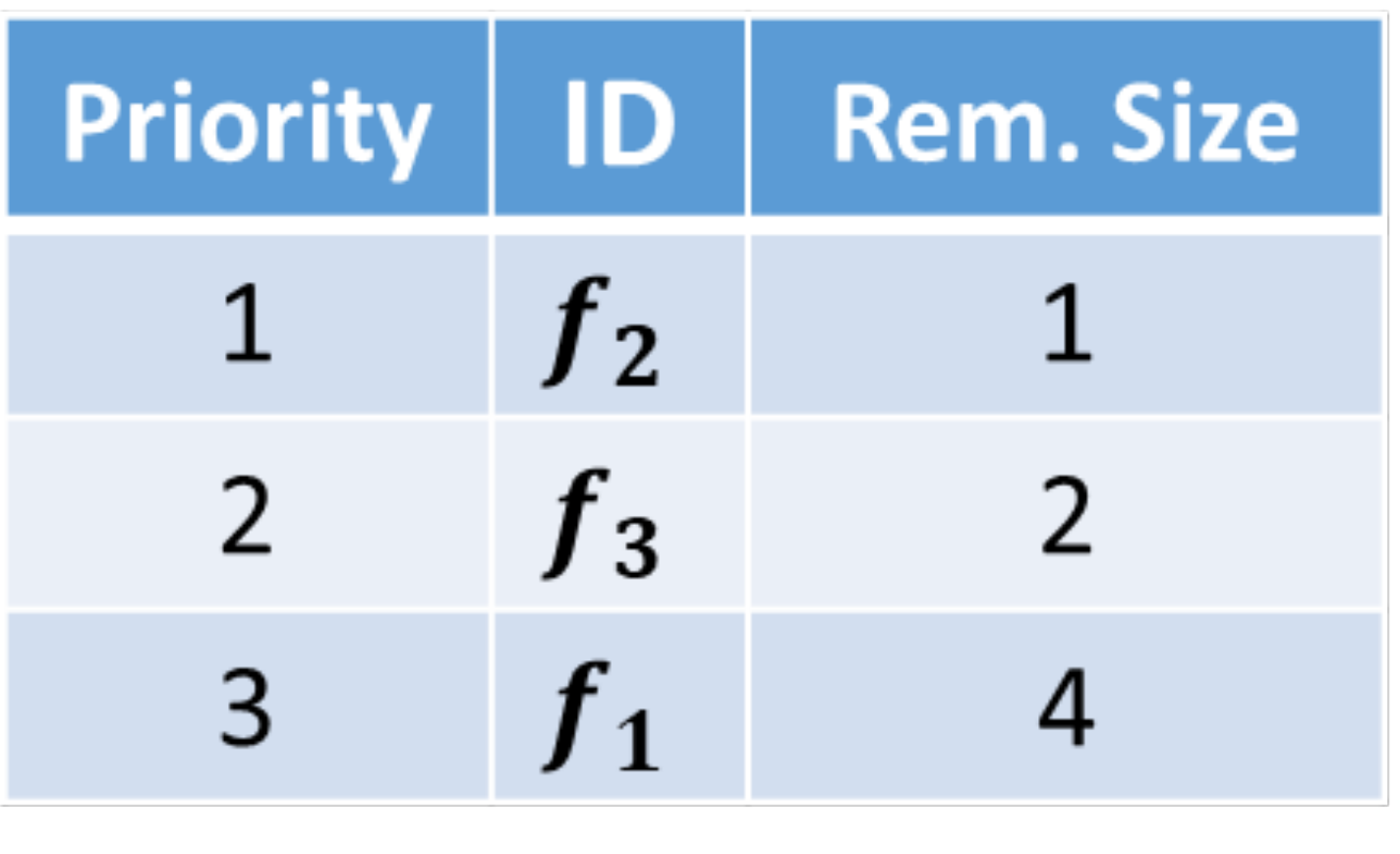}
    \subcaption{Scheduler’s knowledge at time 4 in ideal case}    
    \end{minipage}
    \hfill
    \begin{minipage}[b]{0.48\linewidth}
    \includegraphics[width=\linewidth,height=0.6in]{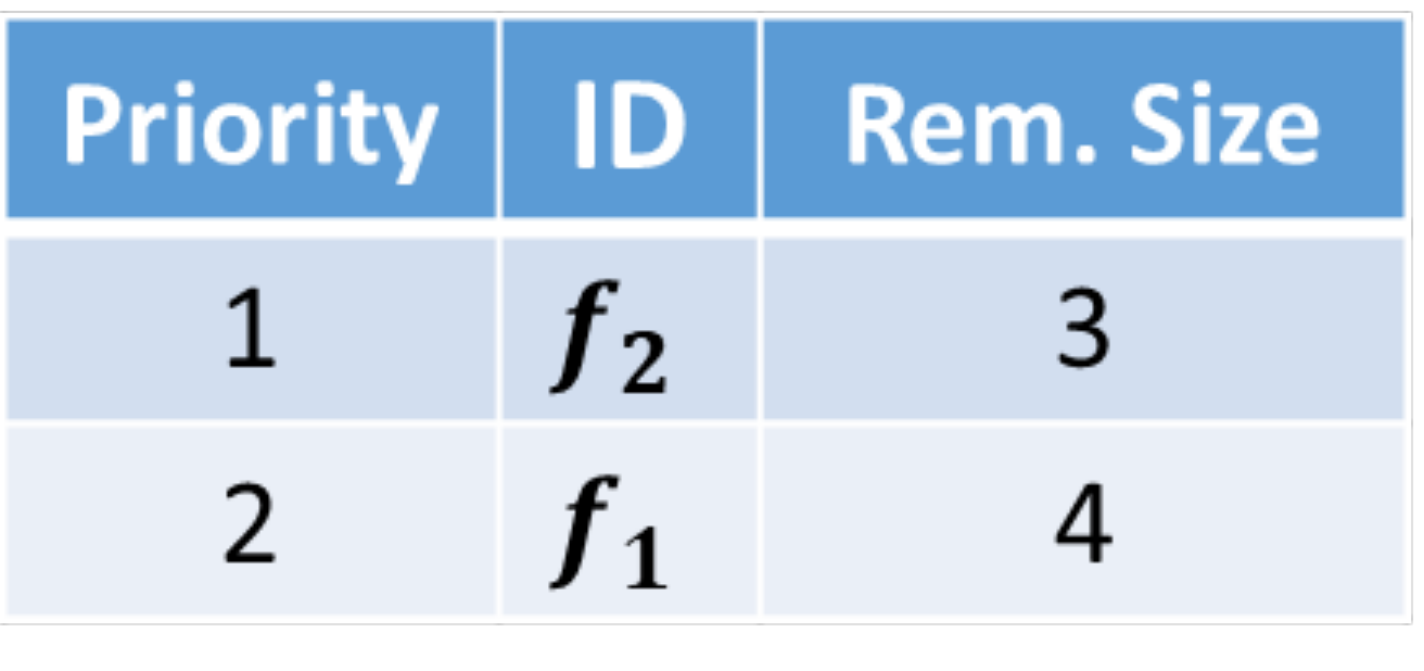}
    \subcaption{Scheduler’s knowledge at time 4.5 in practical case}    
    \end{minipage}
    \hfill
    \begin{minipage}[b]{0.48\linewidth}
    \includegraphics[width=\linewidth,height=0.6in]{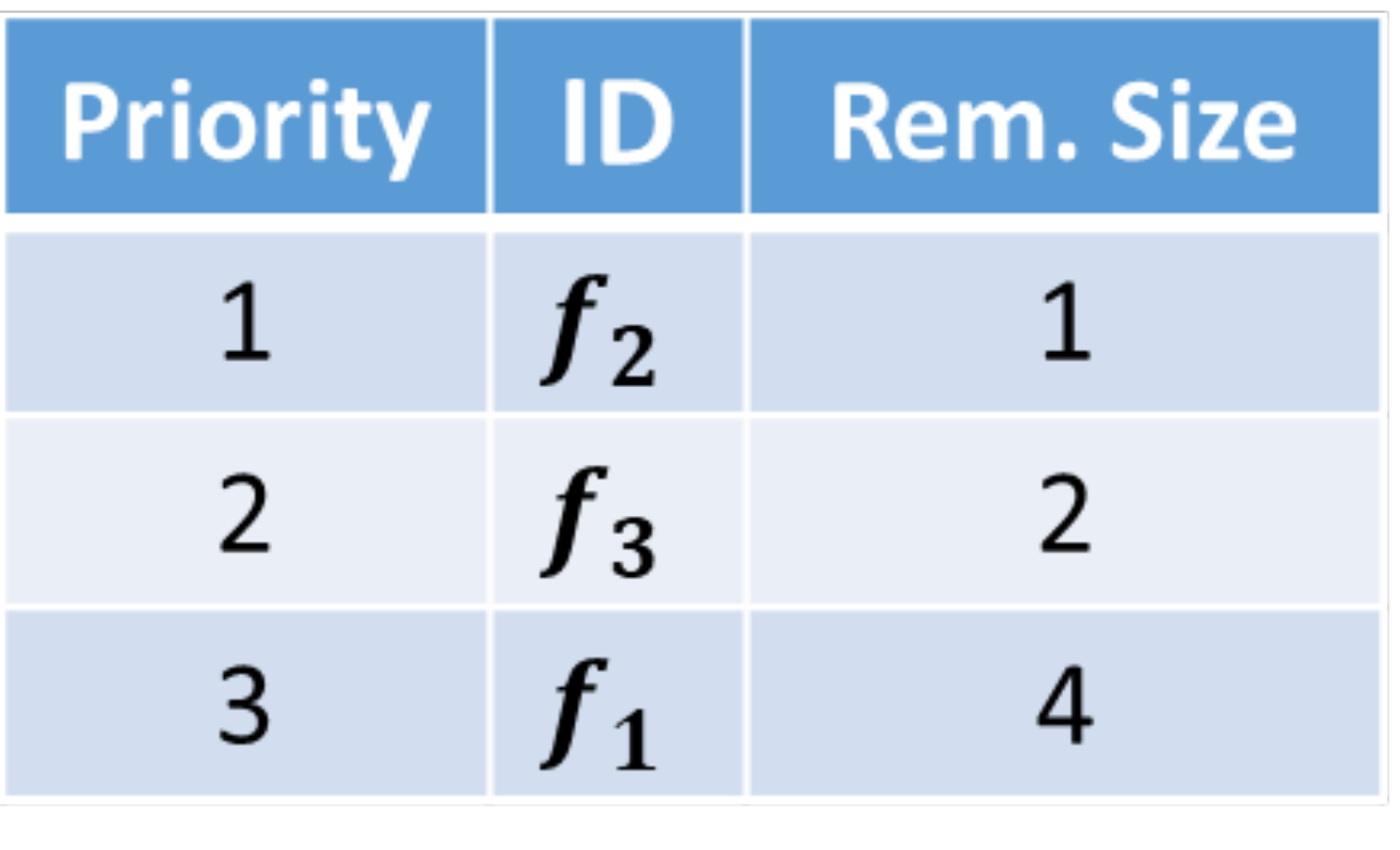}
    \subcaption{Scheduler’s knowledge at time 6.5 in practical case}    
    \end{minipage}
    \hfill
\end{minipage}
\caption{A simple scheduling example}
\label{fig_exm}
\end{figure}

Fig.~\ref{fig_exm} shows a simple example of scheduler’s output in both practical (non-zero scheduling delay) and ideal (zero scheduling delay) scenarios. In this example, three flows named $f_1$, $f_2$, and $f_3$ with 6, 3, and 2 units of data arrive at times 0, 2, and 4, respectively. The total cost of scheduling is 2.5 time units, and service rate (link speed) is 1 unit of data per unit of time. As Fig.~\ref{fig_exm} shows, despite different sizes of the flows, in a practical scenario, all flows are impacted by the scheduling delays.

\subsection{Mean Analysis of SRPT Scheduler Model}
Here, the impact of scheduling delay, $T_{cost}$, on the expected finishing time of flows under SRPT scheduler is formulated. To that end, we first introduce two main propositions stating the impact of scheduling cost on waiting time and on the residence time of flows.

\begin{prop}
Under the presence of scheduling delay, $T_{cost}$, compared to the ideal scenario (when scheduling delay is zero), the waiting time of all flows will be increased exactly by $T_{cost}$.
\end{prop}
\begin{pf}
First, we define the following notations:

$\mathbit{S}_\mathbit{t}^\mathbit{i}$: Sorted list of input flows by the remaining sizes of flows which is known by the scheduler at time t in the ideal scenario ($T_{cost}=0$).

$\mathbit{S}_\mathbit{t}^\mathbit{p}$: Sorted list of input flows by the remaining sizes of flows which is known by the scheduler at time t in the practical scenario ($T_{cost}\neq0$).

$\mathbit{O}_\mathbit{t}^\mathbit{i}$: The flow that is receiving service at time t in the ideal scenario.

$\mathbit{O}_\mathbit{t}^\mathbit{p}$: The flow that is receiving service at time t in the practical scenario.

SRPT scheduler selects the flow with the shortest remaining size (highest priority) among the flows that are already in its flows’ list and serves that flow. Clearly, at any given time t, scheduler’s output only depends on the ordered list of the flows by their remaining sizes. In other words, if for two given times, $t_1$ and $t_2$, we have same list of input flows with the same remaining sizes ($S_{t_1}^i=S_{t_2}^i$), then we should have the same scheduler’s output ($O_{t_1}^i=O_{t_2}^i$). 

In a practical scenario, $T_{cost}$ can be divided into two portions: 
1) sum of network delay from end-hosts to the scheduler and computational delay required to update/generate the current ordered list of flows in the scheduler ($T_1=T_g+T_c$) and 2) network delay of sending the result back ($T_2=T_r$). Therefore, when a new flow comes at time t, it takes $T_1$ timeslots to have an update ordered list of flows calculated by the scheduler. So, compared to the ideal scenario, ordered list of flows is always updated with $T_1$ time delay. Therefore:

\begin{align}
&\forall t\ S_t^i=S_{t+T_1}^p
\label{eq_3a}
\end{align}

Also, in practical scenario, it takes $T_2$ timeslots to send the response back (stop the current lower priority flow receiving service and give service to the new flow if it is the highest priority one). Considering that, in practical scenario, scheduler’s output is always applied with $T_2$ time delay compared to the ideal scenario. In other words, the flow that should be served always receives its service $T_2$ timeslots after when it could ideally be served. So, we have: 

\begin{align}
&\forall t_1,t_2\ \left(S_{t_1}^i=S_{t_2}^p\rightarrow O_{t_1}^i=O_{t_2+T_2}^p\right)
\label{eq_3b}
\end{align}

So, using Eq.~\ref{eq_3a} and~\ref{eq_3b} we have: 

\begin{align}
&\forall t\ O_t^i=O_{\left(t+T_1\right)+T_2}^p=O_{t+T_{cost}}^p
\label{eq_3c}
\end{align}

Now we show that for any new flow we have $w\prime=w+T_{cost}$ where w and $w^\prime$represent waiting time of the new flow in ideal and in practical cases respectively. Suppose not. So, at least there is a flow, $f$, for which we have $w^\prime\neq w+T_{cost}$. Waiting time of a flow is the time from when it first arrives to when it receives service for the first time. So, assume that f arrives at time $t_0$ and receives service for the first time at $t_1$ in ideal case and at $t_1^\prime$ in practical case. Therefore, we have:

\begin{align}
&{(t}_1^\prime-t_0)\neq{(t}_1-t_0)+T_{cost}\Rightarrow t_1^\prime\neq t_1+T_{cost}
\label{eq_3d}
\end{align}

Since $t_1$ and $t_1^\prime$ are the first times that f receives service respectively in ideal and practical scenarios, we should have:

\begin{align}
&\left(\forall t<t_1\ O_t^i\neq f\right) \text{ and } \left(O_{t_1}^i=f\right)
\label{eq_3e}
\end{align}

\begin{align}
& \left(\forall t<{t^\prime}_1\ O_t^p\neq f\right) \text{ and } \left(O_{t_1^\prime}^p=f\right)
\label{eq_3f}
\end{align}

Replacing t with $t_1$ in Eq.~\ref{eq_3c} and using Eq. ~\ref{eq_3e} lead to:

\begin{align}
&O_{t_1}^i=O_{t_1+T_{cost}}^p=f
\label{eq_3g}
\end{align}

However, Eq.~\ref{eq_3f} shows that for all $t<{t^\prime}_1$ we should have $O_t^p\neq f$. Therefore, to meet~\ref{eq_3g}, we should have:

\begin{align}
&t_1+T_{cost}\geq t_1^\prime
\label{eq_3h}
\end{align}

On the other hand, replacing $t$ with ${t'}_1-T_{cost}$ in Eq.~\ref{eq_3c} and using Eq.~\ref{eq_3f} lead to:

\begin{align}
&O_{t_1^\prime-T_{cost}}^i=O_{t_1^\prime}^p=f
\label{eq_3i}
\end{align}

However, Eq.~\ref{eq_3e} shows that for all $t<t_1$ we should have $O_t^p\neq f$. Therefore, to meet Eq.~\ref{eq_3i}, we should have:

\begin{align}
&{t^\prime}_1-T_{cost}\geq t_1
\label{eq_3j}
\end{align}

Now, considering Eq.~\ref{eq_3h} and Eq.~\ref{eq_3j}, we have ${t^\prime}_1=t_1+T_{cost}$ which clearly contradicts Eq.~\ref{eq_3d} and completes the proof.
\qed
\end{pf}

\begin{prop}
The residence time of each flow when there is scheduling delay, $T_{cost}$, is the same as the residence time of the same flow when there is no scheduling delay.
\end{prop}

\begin{pf}
Proposition 1 shows that for any $f$ which receives service for the first time at $t_1$ in ideal case and at $t_1^\prime$ in practical case we have ${t^\prime}_1=t_1+T_{cost}$. Using the same reasoning, it can be shown that when $t_2$ and ${t^\prime}_2$ indicate the finishing time of the f, respectively in ideal and practical cases, we have ${t^\prime}_2=t_2+T_{cost}$. Therefore, residence time of f will remain the same in ideal ($r=t_2-t_1$) and practical ($r=t_2^\prime-t_1^\prime$) cases.
\qed
\end{pf}

As a result of Propositions 1 and 2, considering the scheduling delay, $T_{cost}$, we can calculate the expected completion time for a flow of size x in an M/G/1/SRPT queue, $E\left[T\left(x\right)\right]_{SRPT}$, the expected waiting time of the flow, $E\left[W\left(x\right)\right]_{SRPT}$, and the expected residence time of the flow $E\left[R\left(x\right)\right]_{SRPT}$ as follows (through the rest of this paper we use subscript SRPT-Ideal and SRPT to denote SRPT with zero scheduling delay and SRPT with non-zero scheduling delay, respectively, unless otherwise stated):

\begin{align}
&E\left[W\left(x\right)\right]_{SRPT}=E\left[W\left(x\right)\right]_{SRPT-Ideal}\ +T_{cost}\label{eq_4a}\\
&E\left[R\left(x\right)\right]_{SRPT}=E\left[R\left(x\right)\right]_{SRPT-Ideal}\label{eq_4b}\\
&E[T{\left(x\right)]}_{SRPT}=\left[\frac{\lambda{}\left(m_2\left(x\right)+x^2\left(1-F\left(x\right)\right)\right)}{2{\left(1-\rho{}\left(x\right)\right)}^2}+T_{cost}\right]+\int_0^x\frac{dt}{1-\rho{}\left(t\right)}\label{eq_4c}
\end{align}

\subsection{When No-Scheduling Beats the Best Known Scheduling}
SRPT benefits small flows by prioritizing them over bigger ones and preempting service of the bigger ones to serve these high-priority small flows. These prioritization and preemptive features minimize the queuing delay experienced by small flows in the network so that SRPT can minimize the overall AFCT. However, here, we first show that even when there is no scheduling delay, SRPT does not minimize the completion time of all network’s flows, though it minimizes the average completion time of all flows.

\begin{prop}
For any flow size distribution and any load, $\rho$, less than 1, there always exists some flows for which their expected completion times under FCFS are less than their expected completion times under SRPT-Ideal.
\end{prop}

\begin{pf}
Suppose not. So, for all flows including the largest flow (with size h) completion times are smaller under SRPT-Ideal compared to FCFS. Therefore, considering largest flow, we should have $E\left[R\left(h\right)_{SRPT-Ideal}\right]+E\left[W\left(h\right)_{SRPT-Ideal}\right]<E\left[R\left(h\right)_{FCFS}\right]+E\left[W\left(h\right)_{FCFS}\right]$. Considering that the flow size distribution is bounded, using Eq.~\ref{eq_1a}, ~\ref{eq_1b},~\ref{eq_1c},~\ref{eq_2a},~\ref{eq_2b}, and~\ref{eq_2c} we have: 

\begin{align}
&E\left[R{\left(h\right)}_{SRPT-Ideal}\right]+\frac{\lambda{}m_2\left(h\right)}{2{\left(1-\rho{}\right)}^2}<E\left[R{\left(h\right)}_{FCFS}\right]+\frac{\lambda{}m_2\left(h\right)}{2\left(1-\rho{}\right)}\label{eq_5a}\\
&\frac{\lambda{}{\rho{}.m}_2\left(h\right)}{2{\left(1-\rho{}\right)}^2}<\
E\left[R{\left(h\right)}_{FCFS}\right]-E\left[R{\left(h\right)}_{SRPT-Ideal}\right]\label{eq_5b}
\end{align}

However, since under FCFS flows that are receiving service will not be preempted by other flows, we always have $E\left[R\left(h\right)_{FCFS}\right]\le E\left[R\left(h\right)_{SRPT-Ideal}\right]$. So, Eq.~\ref{eq_5b} becomes $\frac{\lambda{\rho.m}_2\left(h\right)}{2\left(1-\rho\right)^2}<0$. However, this is obviously wrong for $\rho<1$. Therefore, at least, completion time of the largest flow is smaller under FCFS compared to SRPT-Ideal. This contradicts our supposition, and therefore, completes the proof.
\qed
\end{pf}

Proposition 3 shows that SRPT-Ideal minimizes the average completion time of flows by benefiting small flows much more than large ones. In other words, under SRPT-Ideal best performing flows are the small flows. Consequently, we focus on the performance of these best performing small flows. 

Based on that, to investigate the necessity of doing fine-grained scheduling, we introduce a simple two queue system named 2QPlus. In 2QPlus, all network’s flows are classified into two classes: 1st-class includes all flows with sizes less than a threshold (H) and 2nd-class consists of all other flows (this could be done at end-hosts through using available type of service field in IP header (or other available fields such as class of service in VLAN) to identify class of flows’ packets). Then, two queues, which are available in today’s network’s switches, are used. The 1st queue is used to serve all 1st-class flows in FCFS manner, while all other flows will be placed in the 2nd queue and can be served using any scheduling policy such as SRPT or FCFS (Since we are interested in the performance of small flows, scheduling policy of this queue will not impact our analysis). Next, the strict priority mechanism will be used between the two queues so that the 1st-class flows will always be first served. In other words, the new arriving 1st-class flows preempt the service of 2nd-class flows that are currently under service.

Arrival rate of 1st-class flows, $\lambda_H$, is $\lambda_H=\lambda F(H)$ and the load of these flows is $\rho(H)$. We also define $m_1\left(x\right)=\int_{0}^{x}{tf(t)dt}$. So, considering Eq.~\ref{eq_2a},~\ref{eq_2b} and~\ref{eq_2c}, the expected residence time, the expected waiting time, and the expected total completion time of 1st-class flows can be calculated as follows:

\begin{align}
&{E\left[W_H\right]}_{FCFS}=\frac{{\lambda{}}_H\int_0^Ht^2\frac{f\left(t\right)}{F\left(H\right)}dt}{2\left(1-\rho{}\left(H\right)\right)}=\frac{\lambda{}\int_0^Ht^2f\left(t\right)dt}{2\left(1-\rho{}\left(H\right)\right)}=\frac{\lambda{}m_2\left(H\right)}{2\left(1-\rho{}\left(H\right)\right)}\label{eq_6a}\\
&E{\left[R_H\right]}_{FCFS}=\int_0^Hx\frac{f\left(x\right)}{F\left(H\right)}dx=\frac{m_1\left(H\right)}{F\left(H\right)}\label{eq_6b}\\
&E{\left[T_H\right]}_{FCFS}=\frac{\lambda{}m_2\left(H\right)}{2\left(1-\rho{}\left(H\right)\right)}+\frac{m_1\left(H\right)}{F\left(H\right)}\label{eq_6c}
\end{align}

Now, we introduce Theorem~\ref{thm_main} which provides a sufficient condition over H guaranteeing that all 1st-class flows achieve lower AFCT under FCFS compared to SRPT, and consequently shows that 1st-class flows do not require scheduling in DCNs!

\begin{thm}
For any threshold size H that satisfies $E\left[W_H\right]_{FCFS}\le T_{cost}$, the mean completion time of 1st-class flows in 2QPlus system (with sizes less than H) is smaller than or equal to the mean completion time of the same flows using SRPT under any flow size distribution and any load $\rho<1$.
\label{thm_main}
\end{thm}

\begin{pf}
Mean completion time of the flows with sizes less than H under SRPT, $E\left[T_H\right]_{SRPT}$, is: 

\begin{align*}
E{\left[T_H\right]}_{SRPT}=\int_0^HE[T{\left(x\right)]}_{SRPT}\frac{f\left(x\right)}{F\left(H\right)}dx=
\end{align*}
\begin{align}
&=\int_0^HE[W{\left(x\right)]}_{SRPT}\frac{f\left(x\right)}{F\left(H\right)}dx+\int_0^HE[R{\left(x\right)]}_{SRPT}\frac{f\left(x\right)}{F\left(H\right)}dx\label{eq_7a}
\end{align}

Residence time of any flow with size x under any scheduling algorithm is always greater than or equal to the service time required to serve this flow. So, we have ${E\left[R(x)\right]}_{SRPT}\ge x$. Also based on Eq.~\ref{eq_4a}, we have $E[W(x)]_{SRPT}\ge Tcost$. Therefore:

\begin{align}
&E\left[T_H\right]_{SRPT}\geq\int_{0}^{H}{{(T}_{cost}+x)\frac{f(x)}{F\left(H\right)}dx}=T_{cost}+\frac{m_1(H)}{F\left(H\right)}
\label{eq_7b}
\end{align}

We know that $E\left[W_H\right]_{FCFS}=\frac{\lambda m_2\left(H\right)}{2\left(1-\rho\left(H\right)\right)}\le T_{cost}$. By adding $\frac{m_1\left(H\right)}{F\left(H\right)}$ to both sides of this equation and considering (6.c) we have:

\begin{align}
&E\left[T_H\right]_{FCFS}\le T_{cost}+\frac{m_1\left(H\right)}{F\left(H\right)}
\label{eq_7c}
\end{align}

Now, putting Eq.~\ref{eq_7b} and Eq.~\ref{eq_7c} together, we get the following equation which proves Theorem 1:

\begin{align}
E\left[T_H\right]_{FCFS}\le T_{cost}+\frac{m_1\left(H\right)}{F\left(H\right)}\le E\left[T_H\right]_{SRPT}
\end{align}
\qed
\end{pf}

Intuitively, Theorem 1 states that as long as the second moment of the flow size distribution of 1st-class flows remains small, there is no need for scheduling them. This insight becomes important, especially when we consider the fact that DCNs include lots of small flows accounting for just a small portion of total bytes transferred. 

\section{Evaluation}
In this section, we evaluate our analysis in three ways. First, we use two realistic production DCN workloads to find the maximum threshold size H which meets Theorem 1. This value indicates only the sufficient value of H to beat SRPT. So, we go one step further and numerically calculate the exact maximum value of H ($H_{max}$) satisfying $E\left[T_H\right]_{FCFS}\le E\left[T_H\right]_{SRPT}$. In particular, we change various factors including: $T_{cost}$, workload (flow size distribution), and load to check their impact on $H_{max}$. Finally, we simulate 2QPlus and SRPT scheme through extensive flow-level simulations and verify our analysis and results.

\subsection{Flow Size Distributaion in DCNs}
Considering both variability and heavy-tailed nature of DCNs’ traffic and the fact that flow sizes in a production DCN are bounded by minimum and maximum values, we use Bounded-Pareto distribution defined as follows to model the flow size distribution in DCNs:

\begin{align}
&f\left(x\right)=\frac{\alpha k^\alpha}{1-\left(\frac{k}{p}\right)^\alpha}x^{-\alpha-1} & (k\le x\le p,\ 0<\alpha<1)
\label{eq_8}
\end{align}

\begin{figure}[!t]
\center 
\begin{minipage}[t]{0.5\linewidth}
\includegraphics[width=\linewidth,height=1.5in]{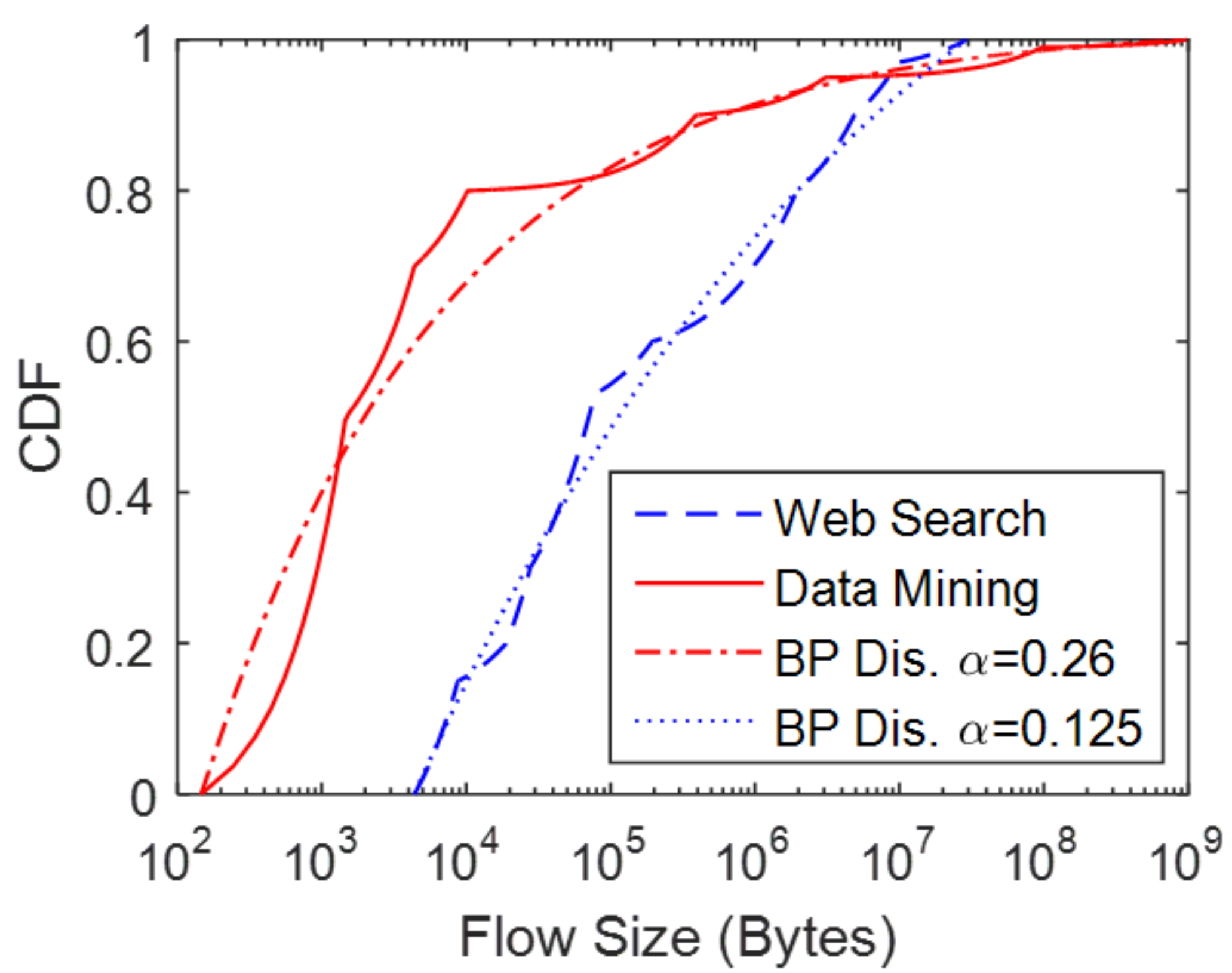}
\subcaption{Flow size distribution}    
\end{minipage}
    \begin{minipage}[b]{0.24\linewidth}
        \center
        \scriptsize
            \begin{tabulary}{.75in}{l|l}
            \hline
             & Value\\
            \hline
            $k$ & 3KB \\
            \hline
            $p$ & 29.2MB \\
            \hline
            $\alpha$ & 0.125 \\
            \hline
            \end{tabulary}
    \subcaption{Parameters for web search’s BP model}    
    \end{minipage}
    \begin{minipage}[b]{0.24\linewidth}
        \center
        \scriptsize
            \begin{tabulary}{0.75in}{l|l}
            \hline
             & Value \\
            \hline
            $k$ & 100B \\
            \hline
            $p$ & 973.34MB \\
            \hline
            $\alpha$ & 0.26 \\
            \hline
            \end{tabulary}
    \subcaption{Parameters for datamining's BP model}    
    \end{minipage}
\caption{Web Search, Data Mining workloads, and their corresponding Bounded-Pareto (BP) models}
\label{fig_web_data}
\end{figure}

In the above formula, the heavy-tailed property of the flow size distribution will be more pronounced when $\alpha$ decreases and vice versa. Also, k and p represent sizes of the smallest and largest flows in the network. To show that how well this distribution models real DCN workloads, we use two realistic DCN workloads obtained from~\cite{dctcp,vl2} and compared them with their corresponding Bounded-Pareto distribution models in Fig.~\ref{fig_web_data}. So, Eq.~\ref{eq_8} provides us with a general way to approximate DCNs’ workloads and a base for the calculations through the rest of the paper.

\subsection{How Much Big Is the Threshold, H, in Today’s DCNs?}
Here, we use web search and data mining workloads (two realistic production DCN workloads which generally cover the pattern of most of the today’s DCNs’ applications) to find the maximum value of H determined by Theorem 1. We use 10Gbps as a typical link capacity used in today’s DCNs~\cite{vl2,jupiter,facebook} to determine service rate of a flow with size x ($x/10Gbps$), and consider $T_{cost}=100\mu$ s which is a typical RTT delay in today’s DCNs~\cite{fastpass,conga}.

\begin{figure}[!t]
\center 
\begin{minipage}[b]{0.48\linewidth}
\includegraphics[width=\linewidth,height=1.3in]{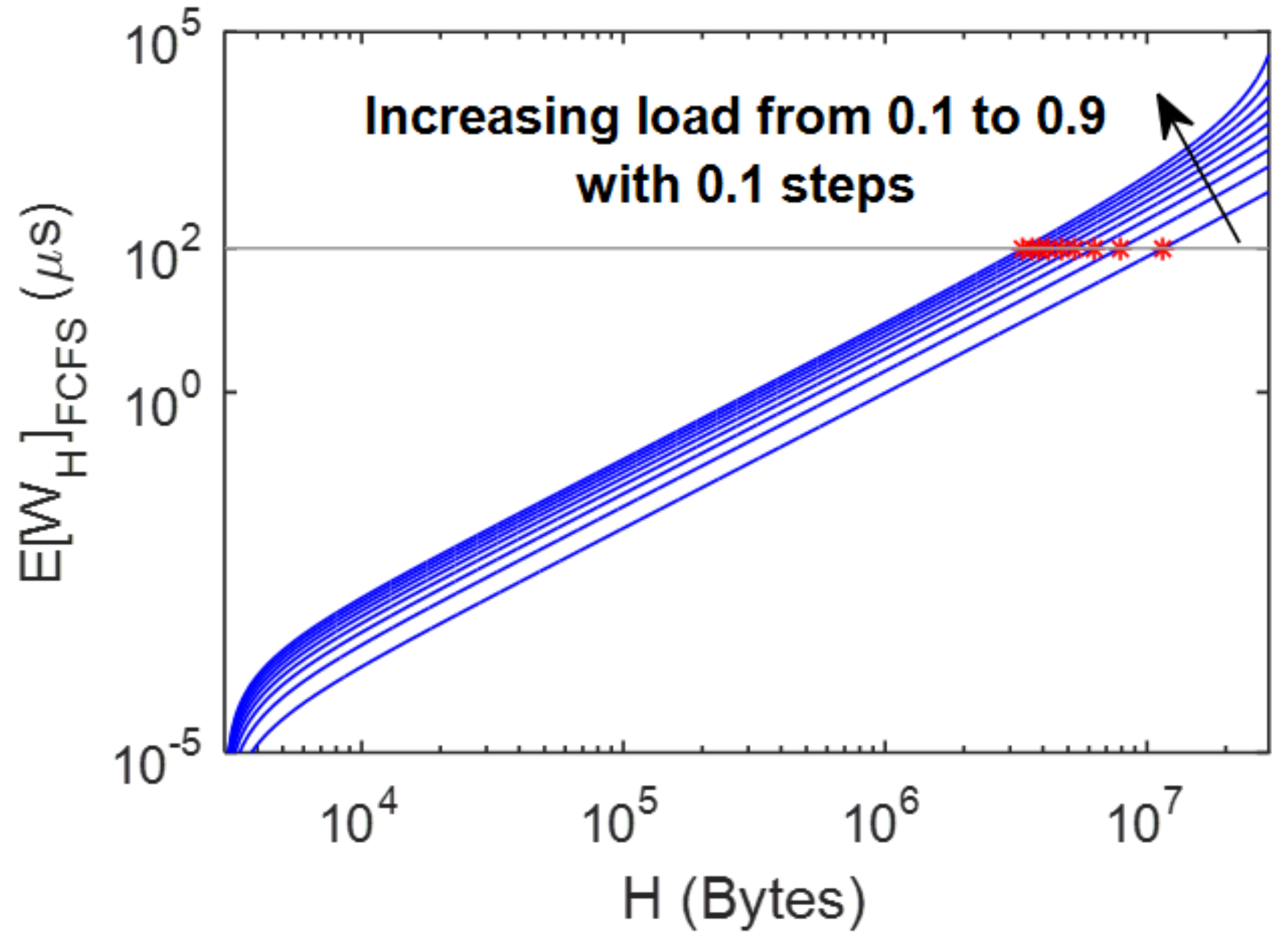}
\subcaption{Web search}    
\end{minipage}
\begin{minipage}[b]{0.48\linewidth}
    \center
    \scriptsize
        \begin{tabulary}{.75in}{l|l|l}
            $\rho$& $H_{max}(MB)$& $F(H_{max})$    \\
            \hline
            0.1      &   11.54         &    0.943        \\
            \hline
            0.2      &   7.89         &    0.918        \\
            \hline
            0.3      &   6.3            &        0.902        \\
            \hline
            0.4      &   5.37         &    0.891        \\
            \hline
            0.5      &   4.74         &    0.882        \\
            \hline
            0.6      &   4.27         &    0.874        \\
            \hline
            0.7      &   3.91         &    0.868        \\
            \hline
            0.8      &   3.63         &    0.862        \\
            \hline
            0.9      &   3.39         &    0.857        \\
            \hline
        \end{tabulary}
\subcaption{$H_{max}$ and ${F(H}_{max})$ for web search workload}    
\end{minipage}
\begin{minipage}[b]{0.48\linewidth}
\includegraphics[width=\linewidth,height=1.3in]{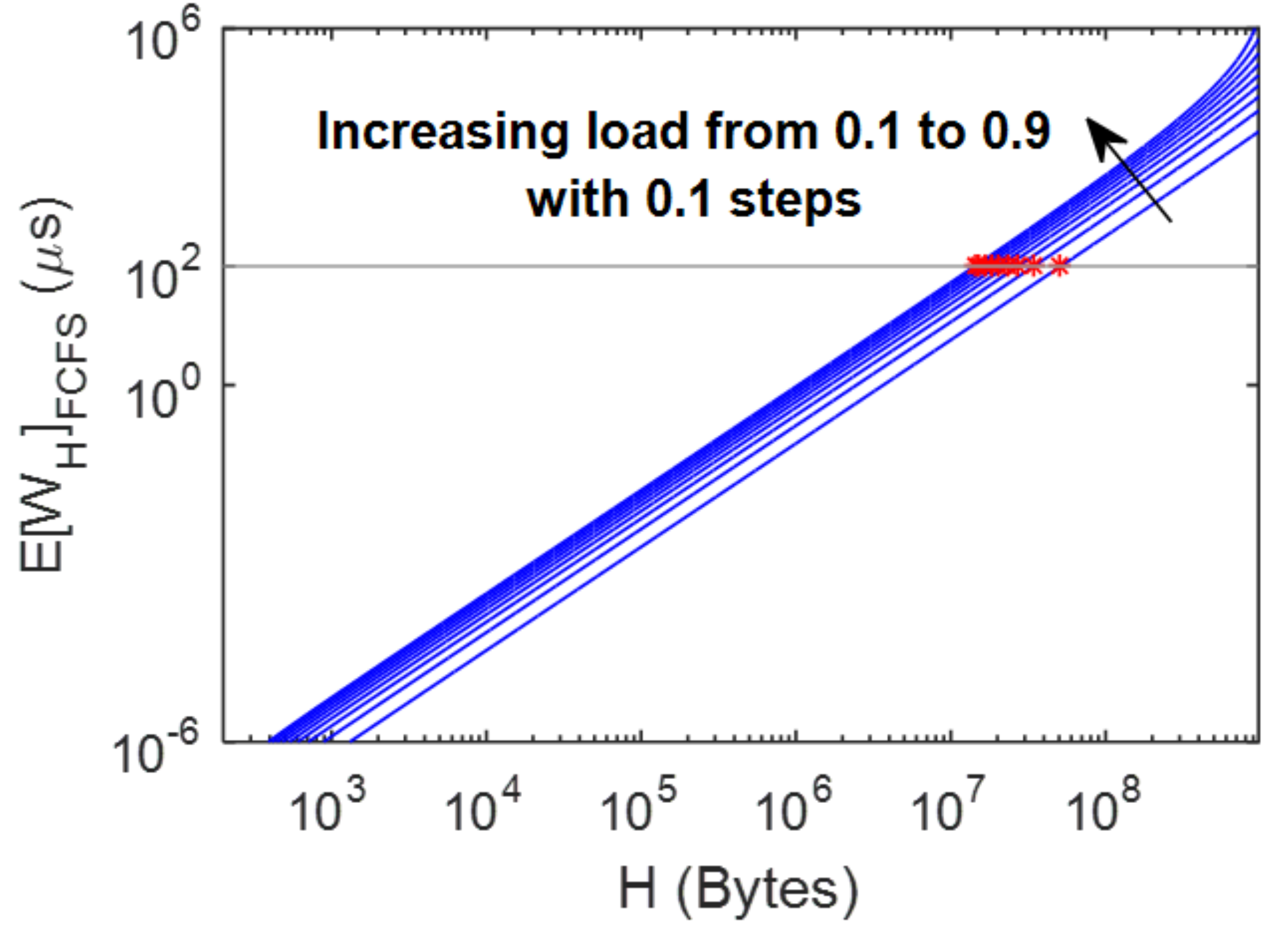}
\subcaption{Data Mining}    
\end{minipage}
\begin{minipage}[b]{0.48\linewidth}
    \center
    \scriptsize
        \begin{tabulary}{.75in}{l|l|l}
            $\rho$ &$    H_{max}(MB)$    &    $F(H_{max})$\\
            \hline
            0.1     &    51.31        &    0.983\\
            \hline
            0.2     &    34.34        &    0.979\\
            \hline
            0.3     &    27.13        &    0.977\\
            \hline
            0.4     &    22.95        &    0.975\\
            \hline
            0.5     &    20.15        &    0.974\\
            \hline
            0.6     &    18.11        &    0.972\\
            \hline
            0.7     &    16.55        &    0.971\\
            \hline
            0.8     &    15.3        &    0.97\\
            \hline
            0.9     &    14.28        &    0.97\\
        \end{tabulary}
        \subcaption{$H_{max}$ and ${F(H}_{max})$ for datamining workload}    
    \end{minipage}
    \caption{$E\left[W_H\right]_{FCFS}$ across different loads (considering $T_{cost}=100\mu$s).}
    \label{fig_wh}
\end{figure}

Fig.~\ref{fig_wh} shows $E\left[W_H\right]_{FCFS}$ versus H for different loads, $\rho$, from 0.1 to 0.9 and for the two mentioned workloads. As Fig.~\ref{fig_wh} illustrates, the maximum values of H ($E\left[W_{H_{max}}\right]_{FCFS}=T_{cost}$) are in the range of $[3.39-11.54MB]$ and [14.28-51.31MB] respectively for web search and data mining workloads. When cumulative distribution function of the flow sizes in each workload is considered, these values of $H_{max}$ show an interesting result that $\approx$86-94\% of web search’s flows and $\approx$97-98\% of data mining’s flows achieve lower AFCT in 2QPlus system without doing any scheduling, compared to when SRPT-the best-known scheduling algorithm-is used.

The threshold H could be changed dynamically based on the load of the network, or simply assigned to be the minimum (or average) value of H (corresponding to the maximum (or average) load in the network). However, since this choice does not impact our results, design of a dynamic mechanism to choose H, will be considered in future work.

\subsection{Numericaly Solved Results}
\label{sec_num}
Theorem 1 only provides a subset of all possible H that can guarantee the 1st-class flows in 2QPlus achive lower AFCT than in SRPT. Here, we would like to explore all possible H that can meet the goal. So, we numerically calculate $E\left[T_H\right]_{SRPT}$ determined by Eq.~\ref{eq_7a} and $E\left[T_H\right]_{FCFS}$ determined by Eq.~\ref{eq_6c} to find all H values meeting the following inequality:

\begin{align}
&\frac{E\left[T_H\right]_{FCFS}}{E\left[T_H\right]_{SRPT}}\le 1
\label{eq_9}
\end{align}

In particular, we numerically examine Eq.~\ref{eq_9} to explore the impact of the various load of the flows, different values of $T_{cost}$, and various DCNs’ workloads on the overall results. Again, we consider a typical 10Gbps link capacity used to serve the flows.

\subsubsection{Impact of Traffic Load}

\begin{figure}[!t]
\center 
\begin{minipage}[b]{0.48\linewidth}
\includegraphics[width=\linewidth,height=1.3in]{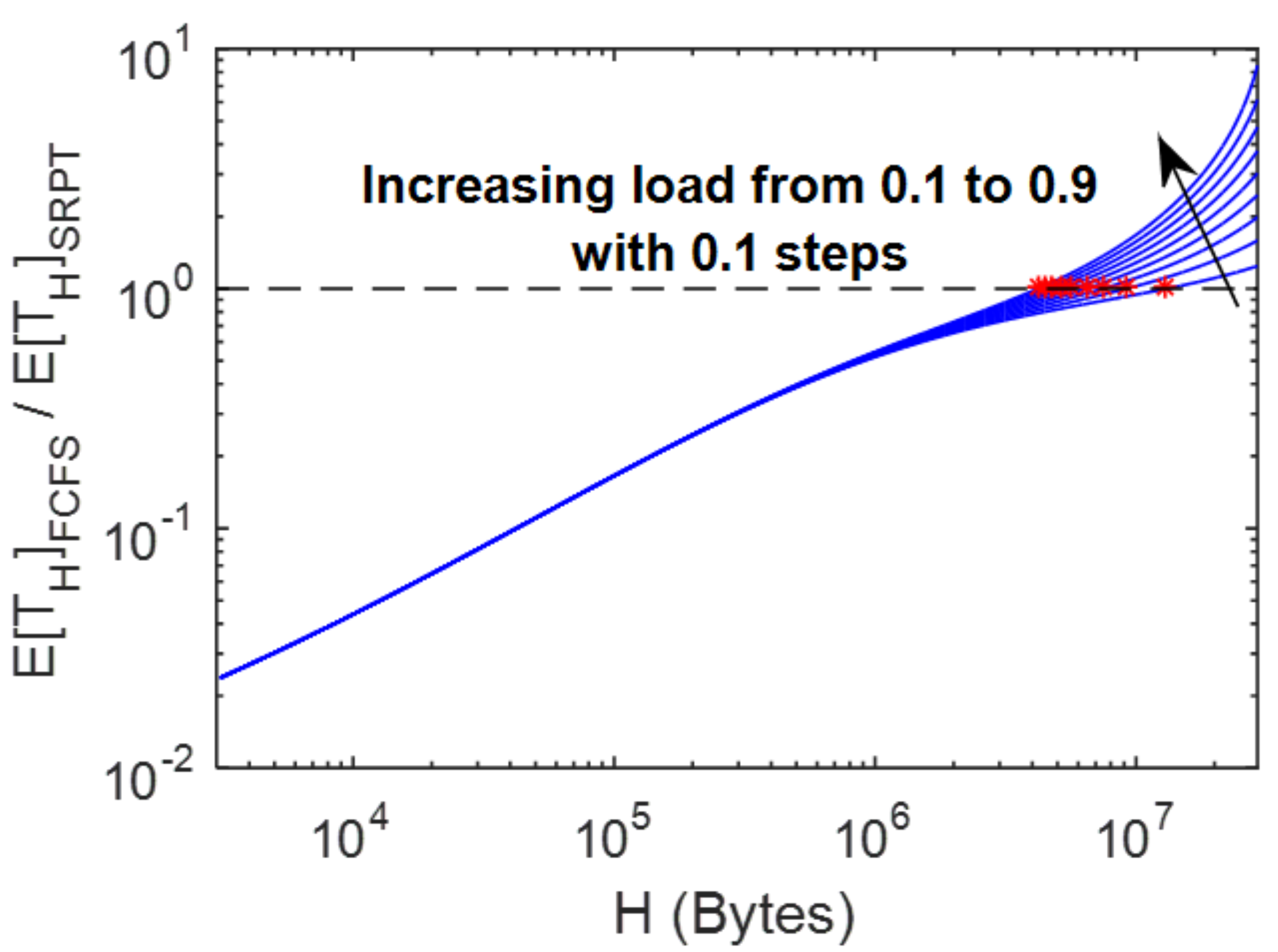}
\subcaption{Web Search}    
\label{fig_th_web_a}
\end{minipage}
\begin{minipage}[b]{0.48\linewidth}
    \center
    \scriptsize
        \begin{tabulary}{.75in}{l|l|l}
            $\rho$&    $H_{max}(MB)$ &    $F(H_{max})$\\
            \hline
            0.1    &    13.1        &    0.951\\
            \hline
            0.2    &    9.18        &    0.928\\
            \hline
            0.3    &    7.45        &    0.914\\
            \hline
            0.4    &    6.42        &    0.904\\
            \hline
            0.5    &    5.72        &    0.895\\
            \hline
            0.6    &    5.2            &    0.889\\
            \hline
            0.7    &    4.8            &    0.883\\
            \hline
            0.8    &    4.47        &    0.878\\
            \hline
            0.9    &    4.2            &    0.873\\
            \hline
        \end{tabulary}
\subcaption{$H_{max}$ and ${F(H}_{max})$ for web search}    
\label{fig_th_web_b}
\end{minipage}
\begin{minipage}[b]{0.48\linewidth}
\includegraphics[width=\linewidth,height=1.3in]{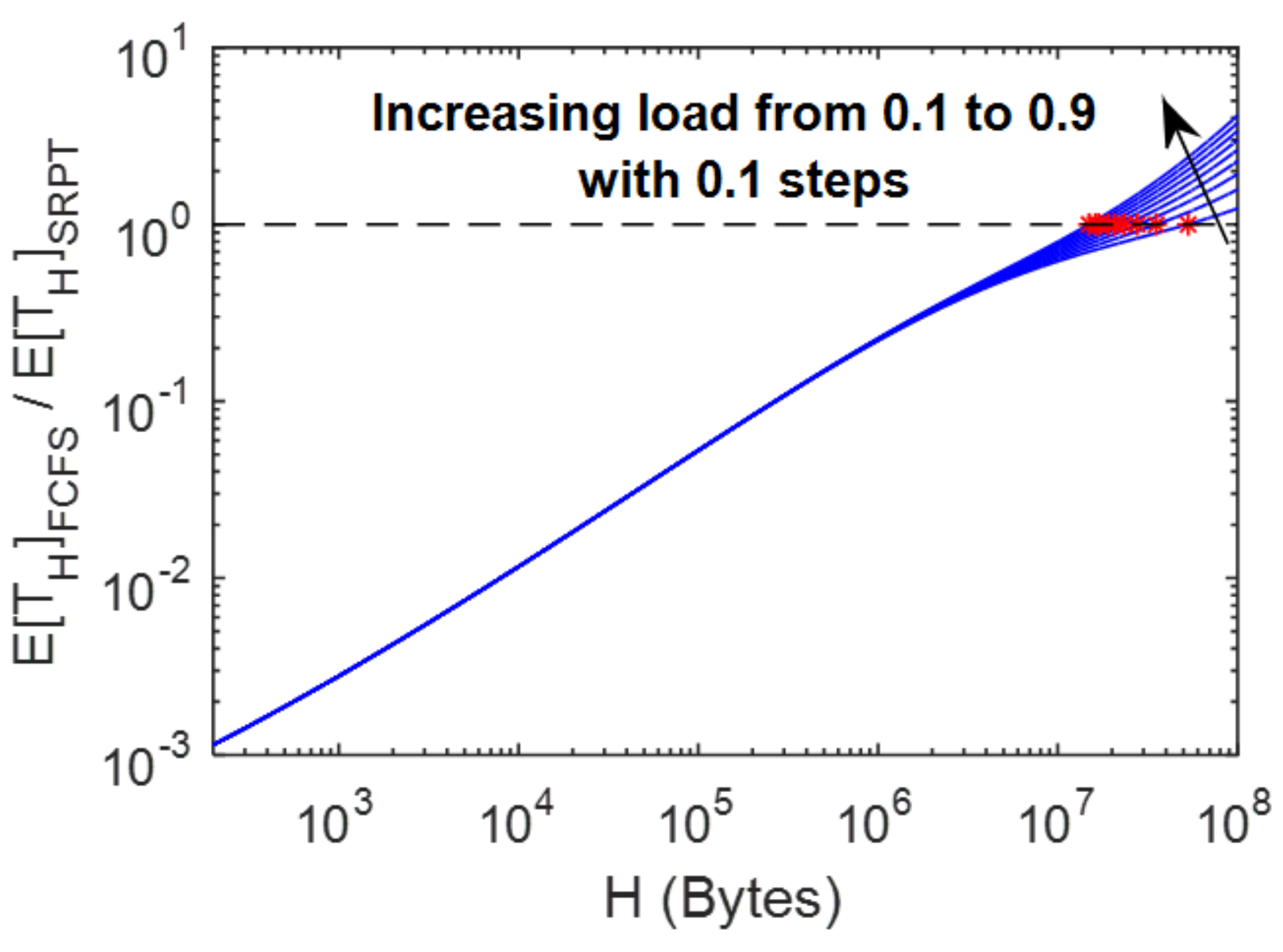}
\subcaption{Datamining}    
\label{fig_th_data_c}
\end{minipage}
\begin{minipage}[b]{0.48\linewidth}
    \center
    \scriptsize
        \begin{tabulary}{.75in}{l|l|l}
            $\rho$&    $H_{max}(MB)$ &    $F(H_{max})$\\
            \hline
            0.1    &        52.7    &        0.983\\
            \hline                           
            0.2    &        35.41    &        0.979\\
            \hline                           
            0.3    &        28.05    &        0.977\\
            \hline                           
            0.4    &        23.77    &        0.975\\
            \hline                           
            0.5    &        20.9    &        0.974\\
            \hline                           
            0.6    &        18.82    &        0.973\\
            \hline                           
            0.7    &        17.21    &        0.972\\
            \hline                           
            0.8    &        15.93    &        0.971\\
            \hline                           
            0.9    &        14.88    &        0.97 \\
            \hline
        \end{tabulary}
        \subcaption{$H_{max}$ and ${F(H}_{max})$ for data mining }    
        \label{fig_th_data_d}
    \end{minipage}
\caption{$E\left[T_H\right]_{FCFS}/E\left[T_H\right]_{SRPT}$ across different loads.}
\label{fig_th}
\end{figure}

To understand the effects of traffic load on the overall results, we fix $T_{cost}$ to 100$\mu$ s, and change the total load, $\rho$, from 0.1 to 0.9 with 0.1 steps. Values of ${E\left[T_H\right]_{FCFS}}/{E\left[T_H\right]_{SRPT}}$ versus H for web search and data mining workloads across different loads are shown in Fig.~\ref{fig_th_web_a} and Fig.~\ref{fig_th_data_c} respectively. 

As Fig 5 declares, for both workloads, till certain values of H, results are similar for various amount of total load. The reason lies on the nature of traffic. All 1st-class flows combined account for a small portion of total bytes transferred. This means that even when the total load is as high as 0.9, the total load of 1st-class flows is still very low. In other words, the inter-arrival of these 1st-class flows is long enough to serve them without having a congestion issue in the network. However, when H increases, the second moment of the flow size distribution for 1st-class flows increases. This causes increase in $E\left[W_H\right]_{FCFS}$ (defined by Eq.~\ref{eq_6a}). Therefore, in the 1st-class, small flows start building up behind bigger ones, and performance drops when H becomes very large.

\subsubsection{Impact of $T_{cost}$}
\begin{figure}[!t]
\center 
\begin{minipage}[b]{0.48\linewidth}
\includegraphics[width=\linewidth,height=1.3in]{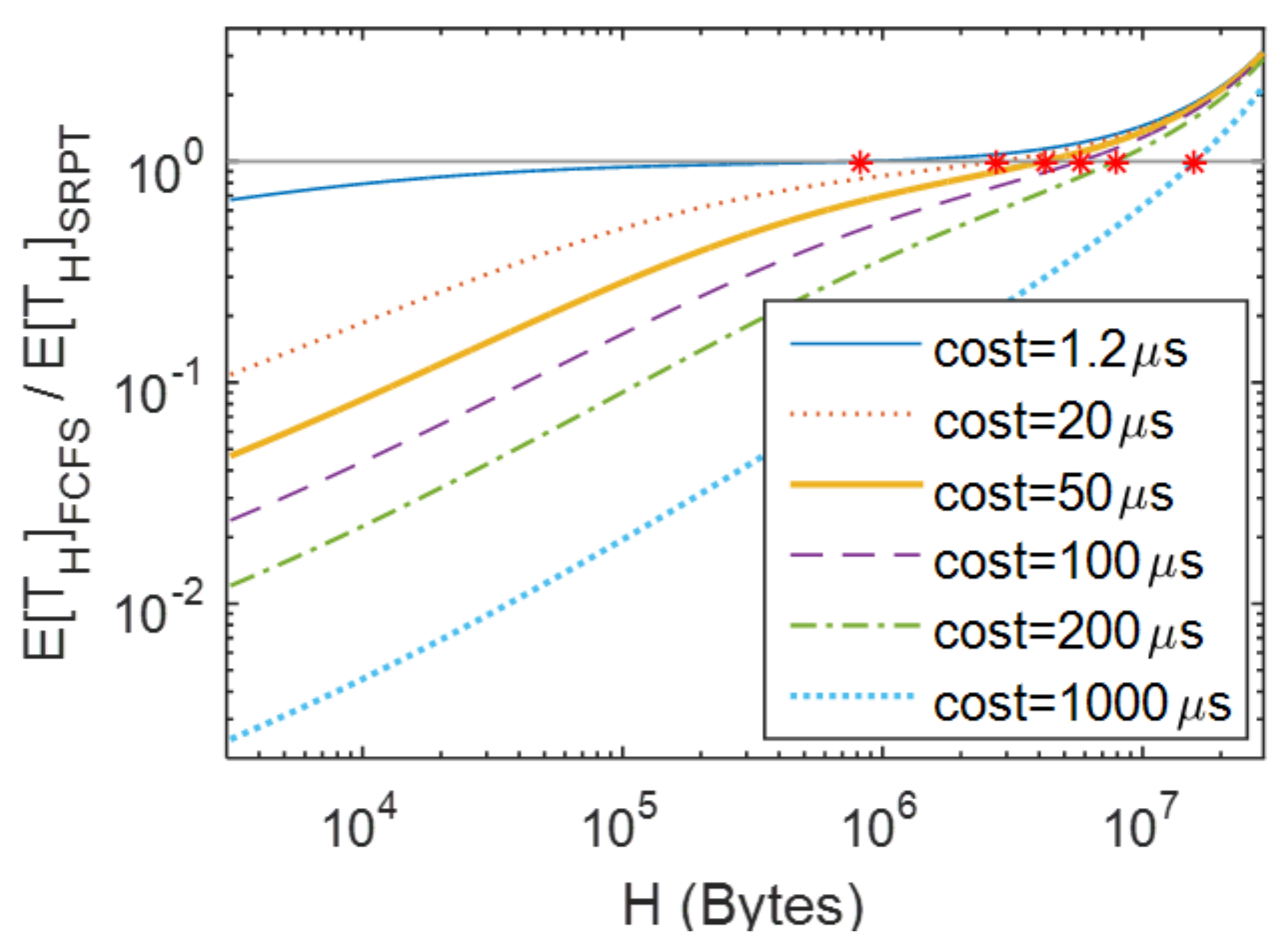}
\subcaption{Web Search}    
\label{fig_th_web_sch_a}
\end{minipage}
\begin{minipage}[b]{0.48\linewidth}
    \center
    \scriptsize
        \begin{tabulary}{.75in}{l|l|l}
            $T_{cost}(\mu s)$ &    $H_{max}(MB)$ &        $F(H_{max})$\\
            \hline
            2.4                &    1.09        &        0.764\\
            \hline
            20                &    2.74        &        0.84\\
            \hline
            50                &    4.16        &        0.872\\
            \hline
            100                &    5.72        &        0.895\\
            \hline
            200                &    7.84        &        0.917\\
            \hline
            1000            &    15.9        &        0.964\\
            \hline
        \end{tabulary}
\subcaption{$H_{max}$ and ${F(H}_{max})$ for web search}    
\label{fig_th_web_sch_b}
\end{minipage}
\begin{minipage}[b]{0.48\linewidth}
\includegraphics[width=\linewidth,height=1.3in]{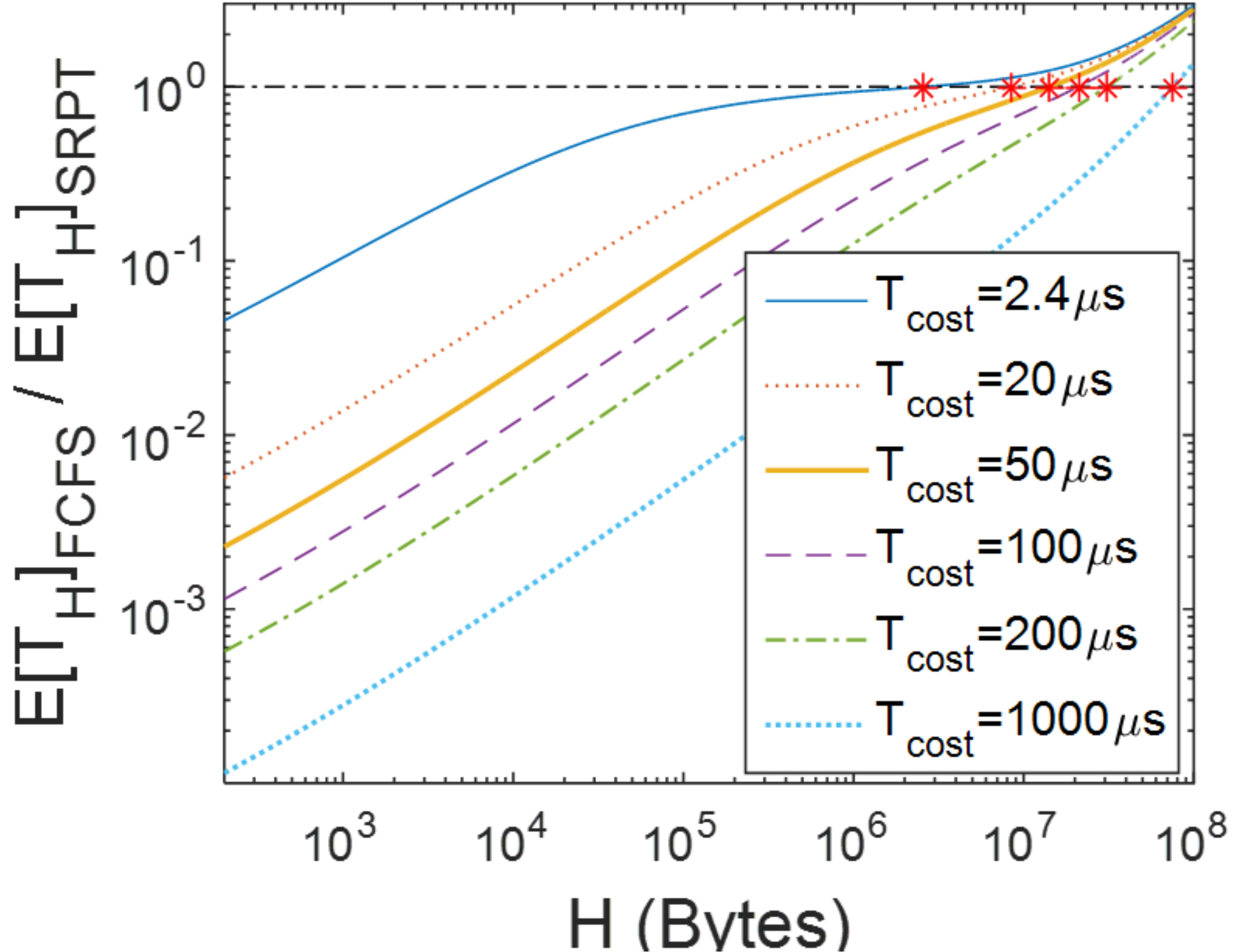}
\subcaption{Datamining}    
\label{fig_th_data_sch_c}
\end{minipage}
\begin{minipage}[b]{0.48\linewidth}
    \center
    \scriptsize
        \begin{tabulary}{.75in}{l|l|l}
            $T_{cost}(\mu s)$ &    $H_{max}(MB)$ &        $F(H_{max})$\\
            \hline
            2.4                &2.59            &    0.944\\
            \hline                                   
            20                &8.48            &    0.963\\
            \hline                                   
            50                &14.17            &    0.969\\
            \hline                                   
            100                &20.9            &    0.974\\
            \hline                                   
            200                &30.81            &    0.978\\
            \hline                                   
            1000            &75.28            &    0.986\\
            \hline
        \end{tabulary}
        \subcaption{$H_{max}$ and ${F(H}_{max})$ for data mining }    
        \label{fig_th_data_sch_d}
    \end{minipage}
\caption{$E\left[T_H\right]_{FCFS}/E\left[T_H\right]_{SRPT}$ across different scheduling delays.}
\label{fig_th_sch}
\end{figure}

Clearly, Theorem 1’s condition depends on the value of $T_{cost}$. So, here, we vary $T_{cost}$ to see its impact on the overall results. In particular, we set load to typical value of 0.5 and vary $T_{cost}$ from as low as $2.4\mu$ s to as high as $1000\mu$ s. $1.2\mu$ s is the transmission time of one packet (1.5KB) over a 10Gbps link. Clearly, RTT (including OS’ stack delays at end-hosts, propagation delay, serialization of a message with sizes more than one packet, and queueing delays at switches) is much higher than $2\times1.2\mu$s~\cite{fastpass,conga,timely}. So, simply, we can say that $T_{cost}$ cannot be smaller than $2.4\mu$s. Fig.~\ref{fig_th_web_sch_a} and Fig.~\ref{fig_th_data_sch_c} show ${E\left[T_H\right]_{FCFS}}/{E\left[T_H\right]_{SRPT}}$ versus H across different $T_{cost}$ values for web search and data mining workloads, respectively.

As expected, increasing $T_{cost}$ increases $H_{max}$, because scheduling cost dominates the waiting time of the 1st-class flows under SRPT. However, even for low values of $T_{cost}$, a big portion of flows still achieve lower AFCT when there is no scheduling. For instance, for $T_{cost}$ as low as $20\mu$s, when the load is 0.5, for the web search workload consisting more big flows and seemingly requiring better scheduling schemes, more than 84\% of the flows still perform better when no scheduling is used to control their access to the network. This number increases to 96\% for the data mining workload which has a broader range of flow sizes and also more small flows compared to the web search workload. Interestingly, for $T_{cost}=2.4\mu$s which is far less from the real RTT delay even when kernel space bypassing techniques is used in special use cases (e.g., check $60\mu$s RTT delay seen after using remote direct memory access (RDMA) and hardware time stamping in~\cite{timely}), $\approx$76\% and $\approx$94\% of all network’s flows perform better under no-scheduling scheme, respectively for search and data mining workloads.

\subsubsection{Impact of Workload}
So far, for the evaluation, we used two realistic DCN workloads. However, there might be still two main concerns about the overall results:

\begin{itemize}
\item What if traffic consists of more small flows?
\item What if a workload consists of more big flows?
\end{itemize}

\begin{figure}[!t]
\center 
\begin{minipage}[t]{0.5\linewidth}
\includegraphics[width=\linewidth,height=1.5in]{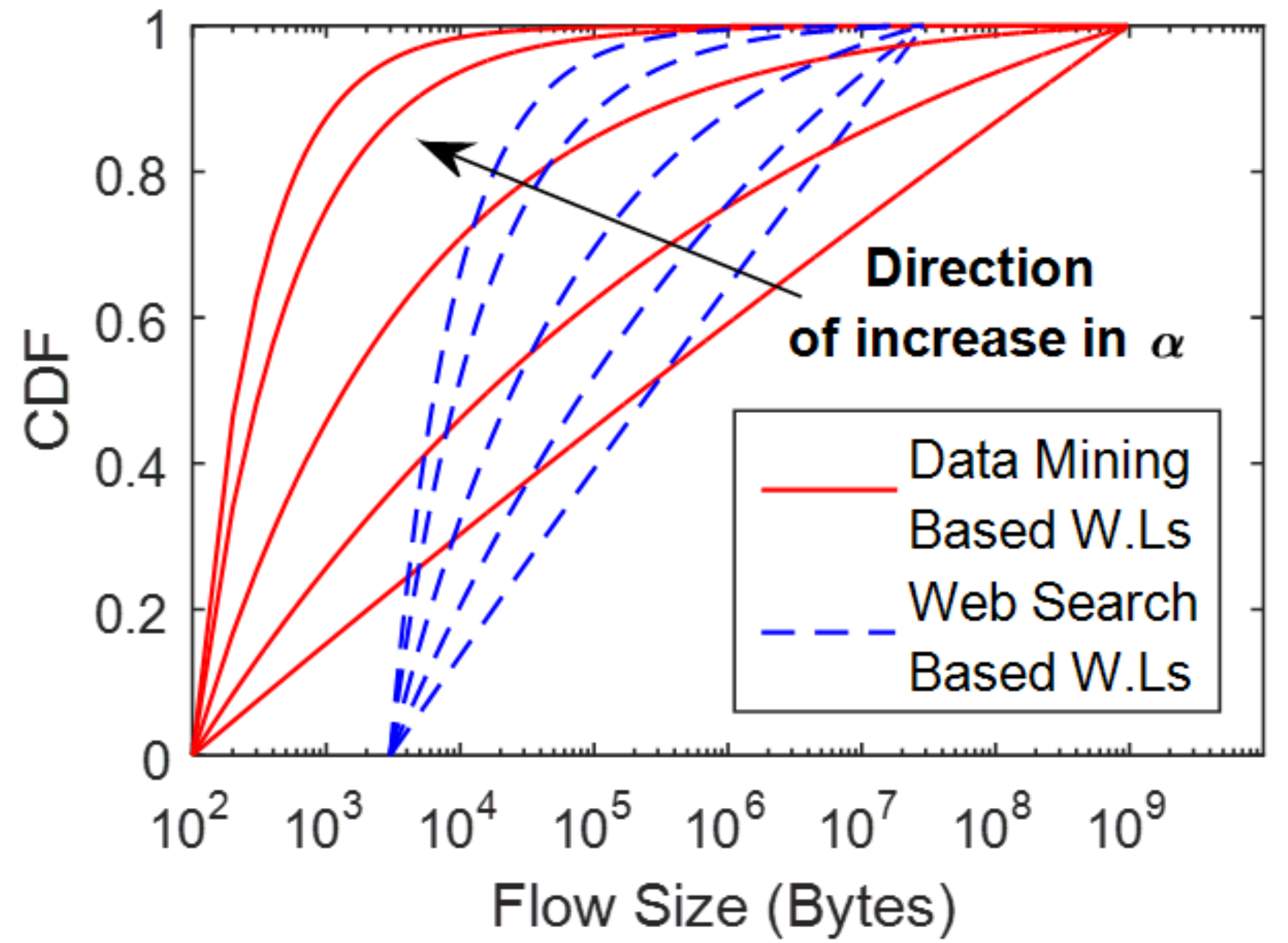}
\subcaption{Flow size distribution}    
\label{fig_synth_a}
\end{minipage}
    \begin{minipage}[b]{0.24\linewidth}
        \center
        \scriptsize
            \begin{tabulary}{.75in}{l|l}
                $\alpha$    &\rot[60]{flows $<$ 100KB}\\
                \hline
                0.01        &45\%\\
                \hline
                0.1        &62.4\%\\
                \hline
                0.26        &84.7\% \\
                \hline
                0.6        &98.5\% \\
                \hline
                0.9        &99.9\% \\
                \hline
            \end{tabulary}
    \subcaption{Datamining based workloads}    
    \label{fig_synth_b}
    \end{minipage}
    \begin{minipage}[b]{0.24\linewidth}
        \center
        \scriptsize
            \begin{tabulary}{0.75in}{l|l}
                $\alpha$    &\rot[60]{flows $<$ 100KB}\\
                \hline
                0.01    &39.3    \%\\
                \hline       
                0.125    &52  \%\\
                \hline       
                0.3        &69.5\%\\
                \hline       
                0.6        &88.2\%\\
                \hline       
                0.9        &95.8\%\\
                \hline
            \end{tabulary}
    \subcaption{Web search based workloads}    
    \label{fig_synth_c}
    \end{minipage}
\caption{Two sets of synthetic workloads used for evaluations, and their percentage of flows smaller than 100KB.}
\label{fig_synth}
\end{figure}

To evaluate the results under these corner cases, we used Bounded-Pareto distribution model defined by Eq.~\ref{eq_8}. As mentioned in section IV-A, the heavy-tailed property of this model is more pronounced when $\alpha$ decreases and vice versa. So, we change $\alpha$ from 0.01 to 0.9 to generate two different sets of synthetic workloads shown in Fig.~\ref{fig_synth}. For the first set (data mining based workloads), minimum and maximum flow sizes of the data mining workload are used, while minimum and maximum sizes of the web search workload are used for the second set (web search based workloads). Percentage of flows that are smaller than 100KB for both sets of workloads is shown in Fig.~\ref{fig_synth_b} and Fig.~\ref{fig_synth_c}. For instance, in the first set, for $\alpha=0.9, 99.9\%$ of flows are smaller than 100KB, while for $\alpha=0.01$, this number is only 45\%. So, these sets provide us with proper workloads to check the two mentioned corner cases. 

\begin{figure}[!t]
\center 
\begin{minipage}[b]{0.48\linewidth}
\includegraphics[width=\linewidth,height=1.3in]{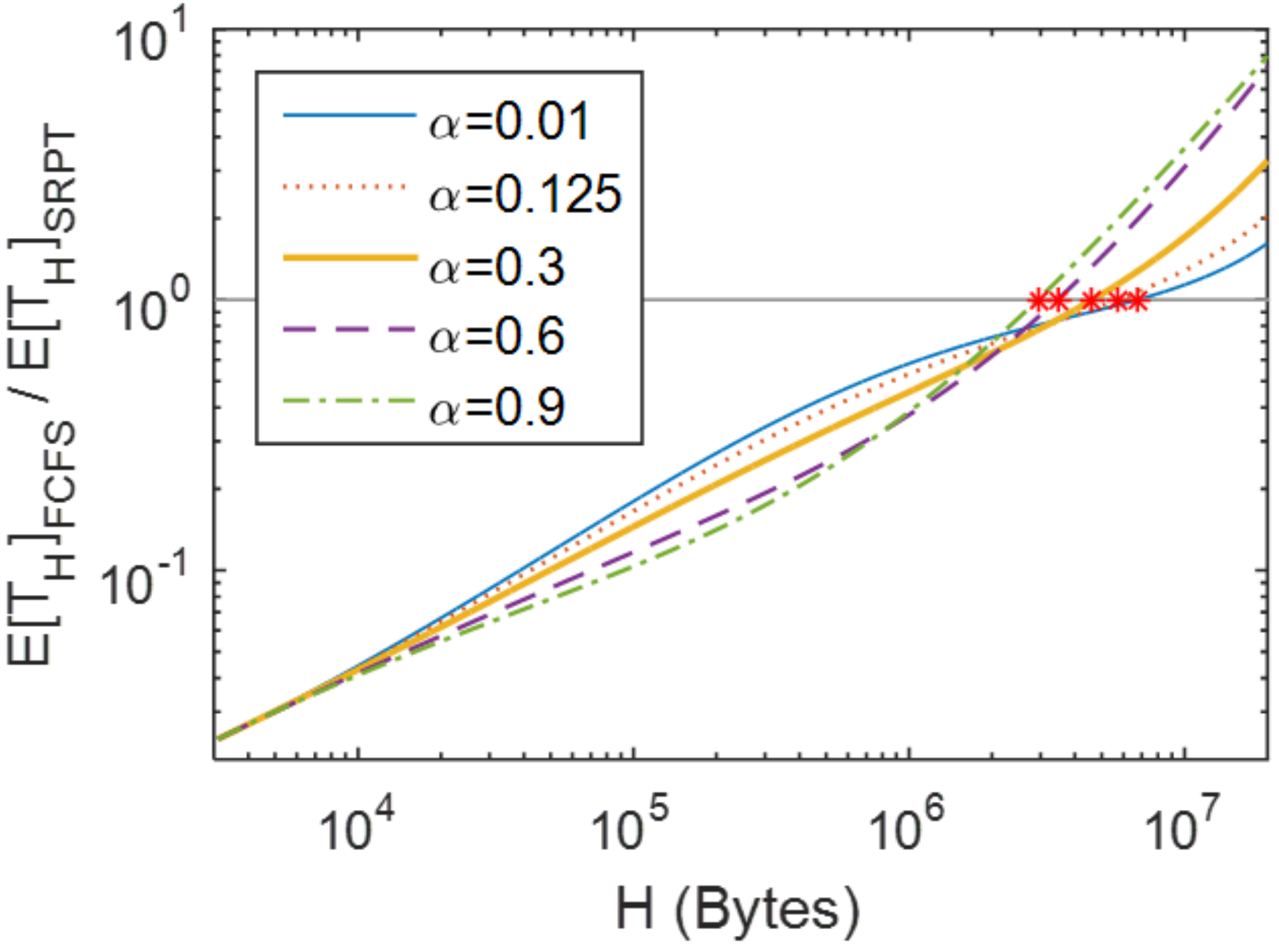}
\subcaption{Web Search}    
\end{minipage}
\begin{minipage}[b]{0.48\linewidth}
    \center
    \scriptsize
        \begin{tabulary}{.75in}{l|l|l}
            $\alpha$    &$H_{max}(MB)$    &$F(H_{max})$\\
            \hline
            0.01    &6.73    &0.847    \\
            \hline                 
            0.125    &5.72    &0.895\\
            \hline                 
            0.3        &4.63    &0.95 \\
            \hline                 
            0.6        &3.45    &0.99 \\
            \hline                 
            0.9        &2.93    &0.999\\
            \hline
        \end{tabulary}
\subcaption{$H_{max}$ and ${F(H}_{max})$ for web search}    
\end{minipage}
\begin{minipage}[b]{0.48\linewidth}
\includegraphics[width=\linewidth,height=1.3in]{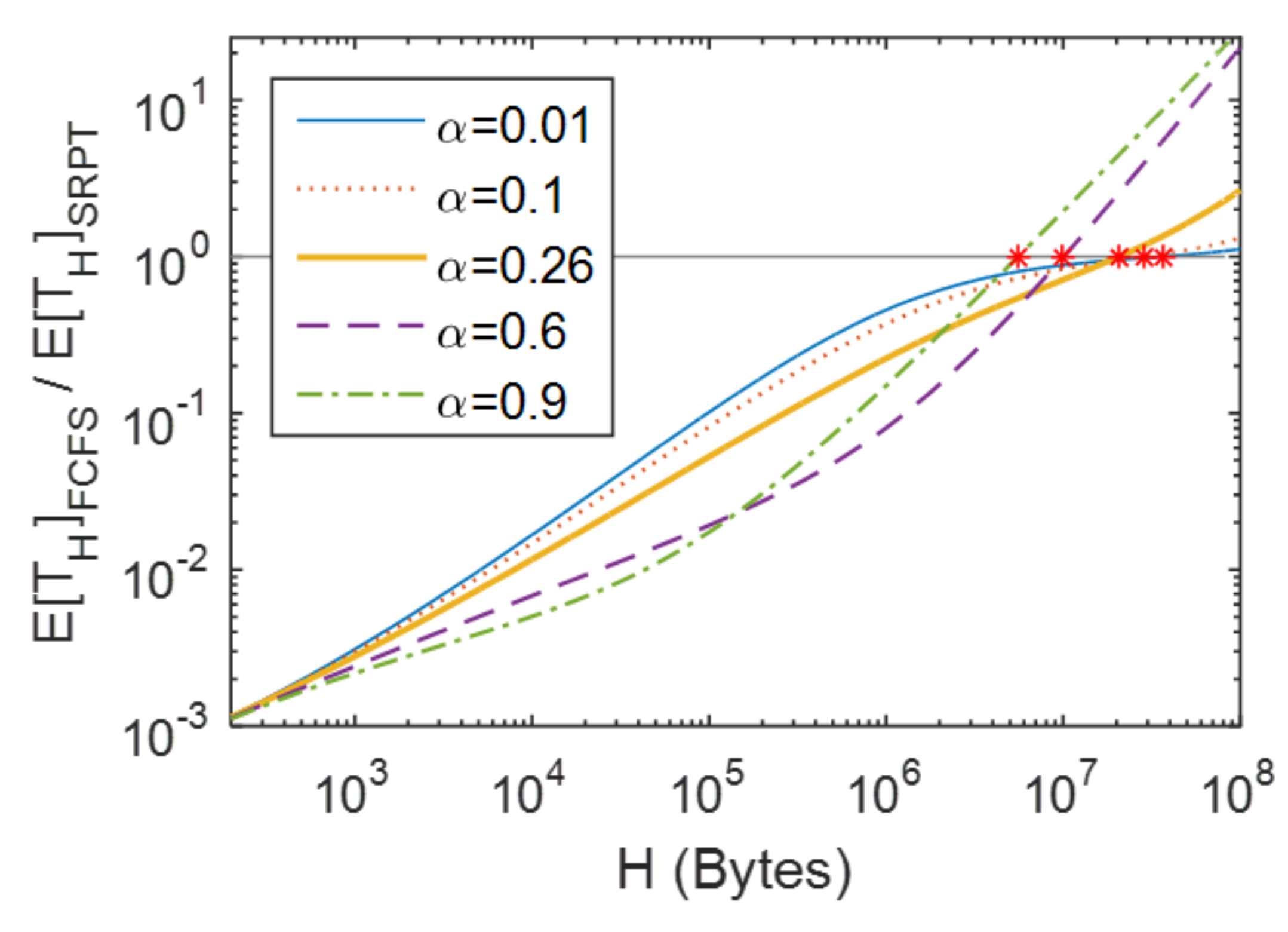}
\subcaption{Datamining}    
\end{minipage}
\begin{minipage}[b]{0.48\linewidth}
    \center
    \scriptsize
        \begin{tabulary}{.75in}{l|l|l}
            $\alpha$    &$H_{max}(MB)$    &$F(H_{max})$\\
            \hline
            0.01    &38.69    &0.813        \\
            \hline                   
            0.1        &29.96    &0.896    \\
            \hline                   
            0.26    &20.9    &0.974    \\
            \hline                   
            0.6        &10.07    &0.9991    \\
            \hline                   
            0.9        &5.56    &0.99995\\
            \hline
        \end{tabulary}
        \subcaption{$H_{max}$ and ${F(H}_{max})$ for data mining }    
    \end{minipage}
\caption{$E\left[T_H\right]_{FCFS}/E\left[T_H\right]_{SRPT}$ across different scheduling delays.}
\label{fig_th_wrk}
\end{figure}

Fig.~\ref{fig_th_wrk} shows the ${E\left[T_H\right]_{FCFS}}/{E\left[T_H\right]_{SRPT}}$ across these different workloads for total load of 0.5 and $T_{cost}=100\mu$s. Increasing heavy-tailed ness of the workload increases the probability of having more concurrent big flows in the network. This affects the results from two aspects. First, decreasing $\alpha$ reduces the number of small flows and consequently, for small values of H, it decreases the congestion among 1st-class flows (with sizes smaller than H) which improves the performance of these flows. Second, for bigger values of H, decreasing $\alpha$ causes more big flows going into the 1st queue. So, the average waiting time of 1st-class flows increases and performance drops. However, despite these impacts, again, most of the flows ($\approx$85\%-99.9\% for web search based workloads and $\approx$81\%-99.9\% in data mining based workloads) perform better under the no-scheduling scheme.

\subsection{Simulations}
To validate our analysis and its results, here, we conduct extensive flow-level simulations. To that end, similar to our analysis, we use the single-queue model of the network and simulate SRPT scheme. Also, we simulate 2QPlus and use SRPT algorithm (with non-zero scheduling delay) to schedule its 2nd-class flows, while 1st-class flows receive service in FCFS manner. We generate more than 40000 flows based on the 2 realistic production DCN workloads used in section IV-C and change load from 0.1 to 0.9. As before, we set $T_{cost}=100\mu$s as a scheduling delay. We consider AFCT as the performance metric and report $AFCT_{2QPlus}/AFCT_{SRPT}$ separately for 1st-class, 2nd-class, and all flows. 

\begin{figure}[!t]
\center 
\begin{minipage}[b]{0.48\linewidth}
\includegraphics[width=\linewidth,height=1.3in]{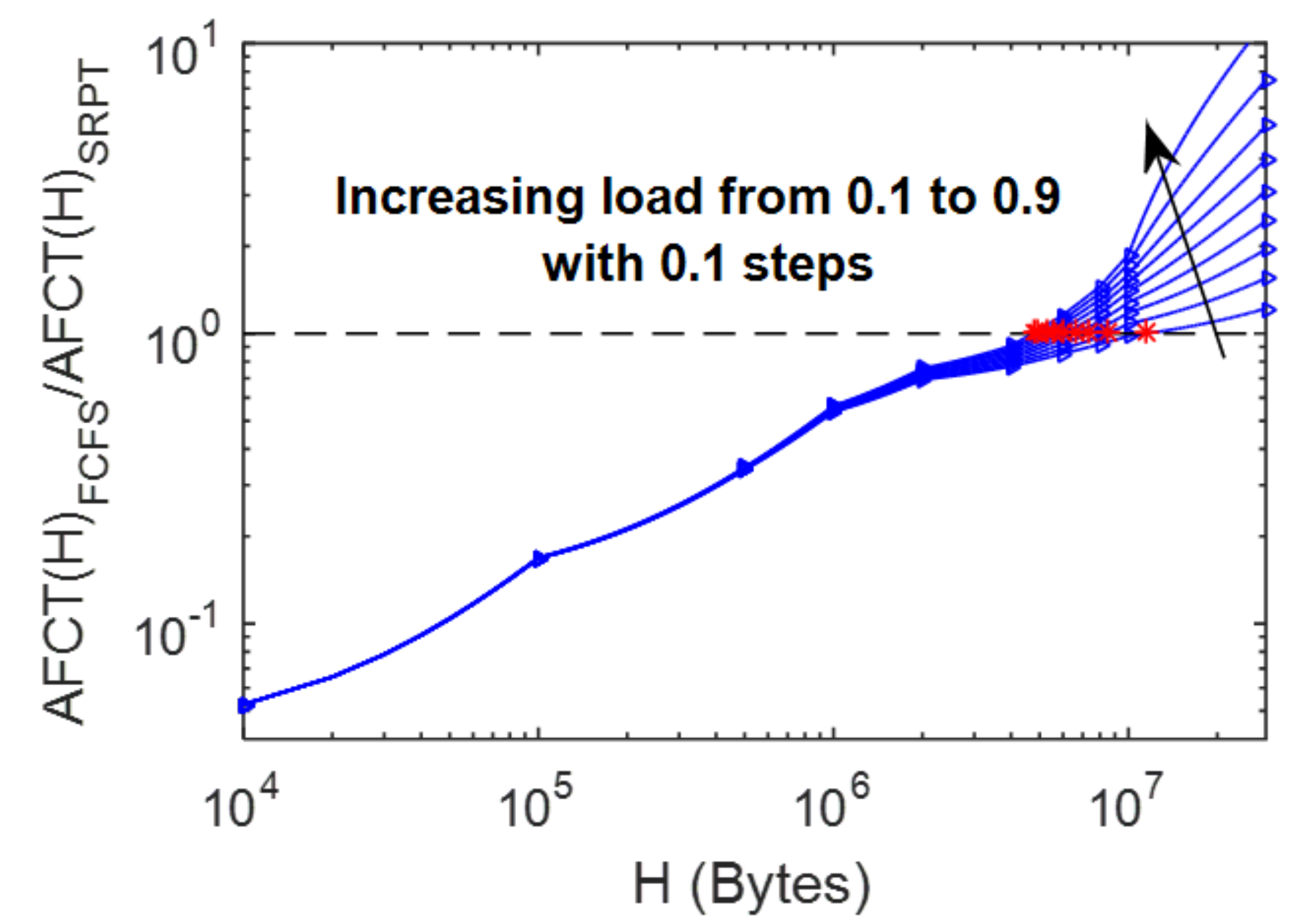}
\subcaption{Web Search}    
\end{minipage}
\begin{minipage}[b]{0.48\linewidth}
    \center
    \scriptsize
        \begin{tabulary}{.75in}{l|l|l}
            $\rho$    & $H_{max}(MB)$    & $F(H_{max})$\\
            \hline
            0.1        &     11.53    &     0.943\\
            \hline                       
            0.2        &     8.48    &     0.923\\
            \hline                       
            0.3        &     7.32    &     0.913\\
            \hline                       
            0.4        &     6.56    &     0.905\\
            \hline                       
            0.5        &     6.02    &     0.899\\
            \hline                       
            0.6        &     5.57    &     0.894\\
            \hline                       
            0.7        &     5.23    &     0.889\\
            \hline                       
            0.8        &     4.97    &     0.885\\
            \hline                       
            0.9        &     4.73    &     0.882\\
            \hline
        \end{tabulary}
\subcaption{$H_{max}$ and ${F(H}_{max})$ for web search}    
\end{minipage}
\begin{minipage}[b]{0.48\linewidth}
\includegraphics[width=\linewidth,height=1.3in]{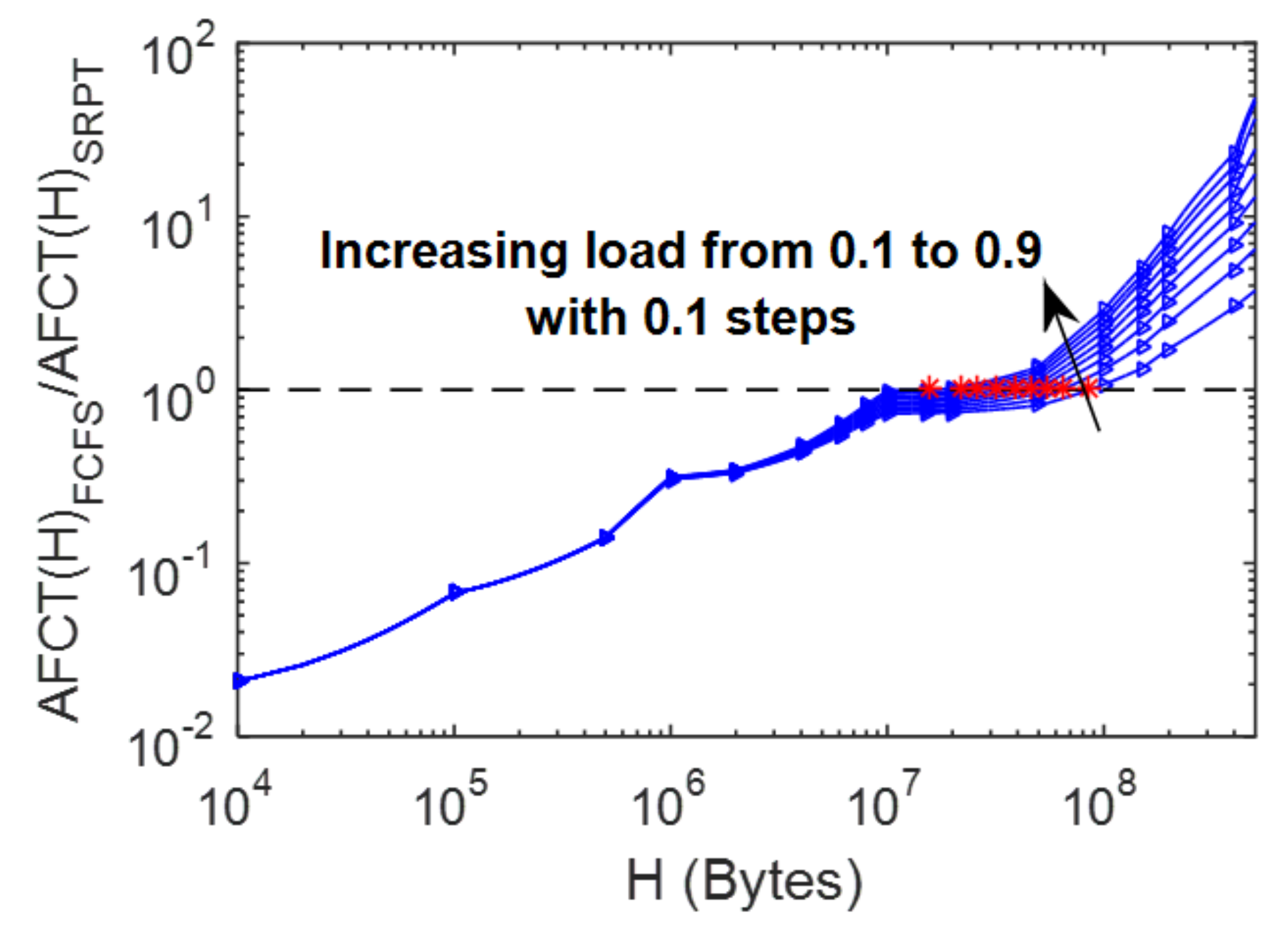}
\subcaption{Datamining}    
\end{minipage}
\begin{minipage}[b]{0.48\linewidth}
    \center
    \scriptsize
        \begin{tabulary}{.75in}{l|l|l}
            $\rho$    & $H_{max}(MB)$    & $F(H_{max})$\\
            \hline              
            0.1        &    85.53    &    0.987\\
            \hline                       
            0.2        &    63.72    &    0.985\\
            \hline                       
            0.3        &    54.08    &    0.983\\
            \hline                       
            0.4        &    46.29    &    0.982\\
            \hline                       
            0.5        &    38.40    &    0.98 \\
            \hline                       
            0.6        &    31.38    &    0.978\\
            \hline                       
            0.7        &    26.18    &    0.976\\
            \hline                       
            0.8        &    22.00    &    0.975\\
            \hline                       
            0.9        &    15.69    &    0.971\\
            \hline
        \end{tabulary}
        \subcaption{$H_{max}$ and ${F(H}_{max})$ for data mining }    
    \end{minipage}
\caption{$E\left[T_H\right]_{FCFS}/E\left[T_H\right]_{SRPT}$ across different scheduling delays for 1st class of flows.}
\label{fig_afct_1}
\end{figure}

Fig.~\ref{fig_afct_1} shows the AFCT of the 1st-class flows which fits very well with the analytical results shown in Fig.~\ref{fig_th}. In more details, $\approx$88\%-94\% and $\approx$97\%-99\% of the flows perform better under the no-scheduling scheme, respectively for the search and data mining workloads.

\begin{figure}[!t]
    \center 
    \begin{minipage}[b]{0.48\linewidth}
        \includegraphics[width=.9\linewidth,height=1.3in]{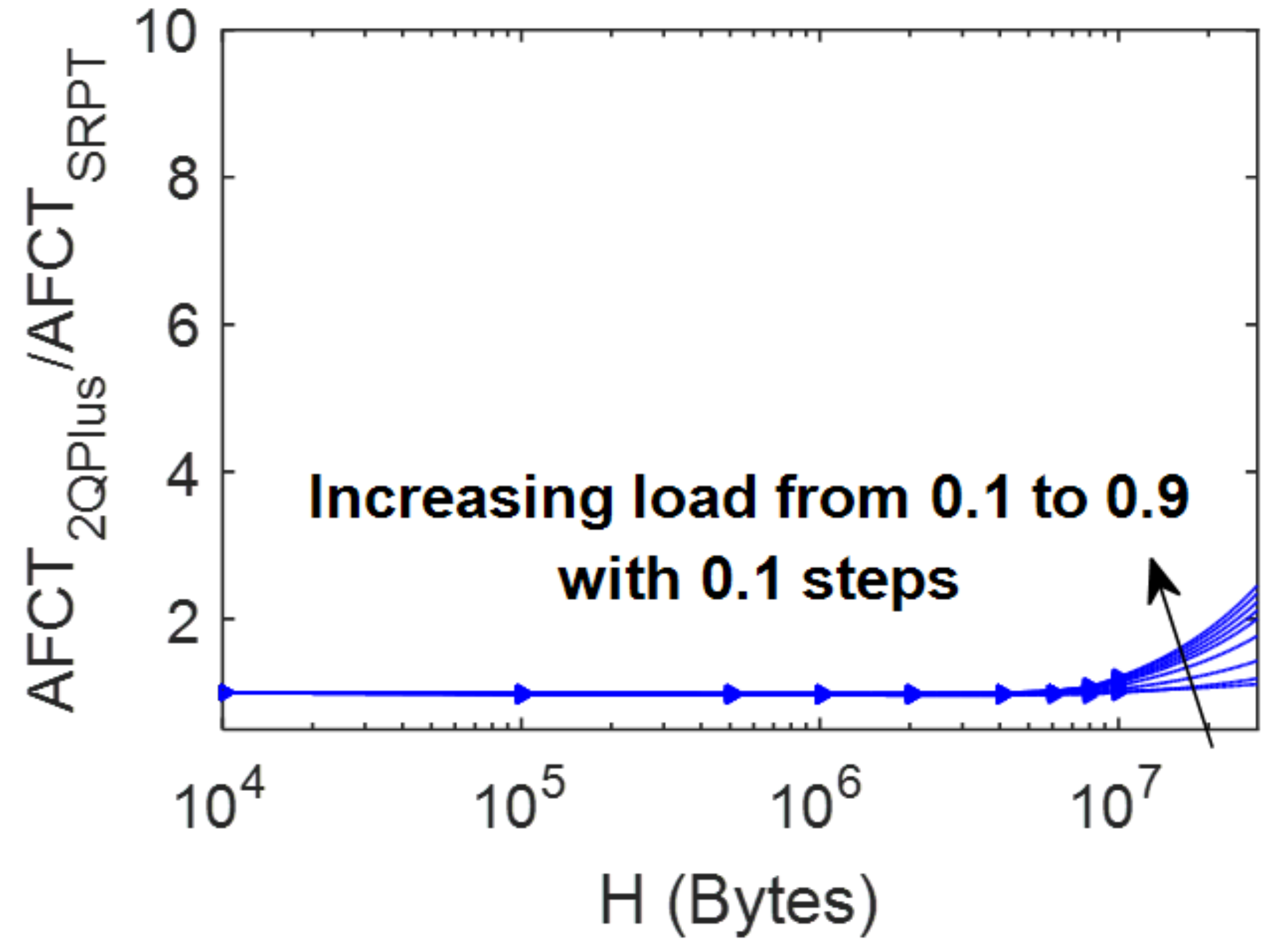}
        \subcaption{Web search’s 2nd-class flows}    
            \label{fig_afct_web_big}
    \end{minipage}
    \hfill
    \begin{minipage}[b]{0.48\linewidth}
        \center
        \includegraphics[width=.9\linewidth,height=1.3in]{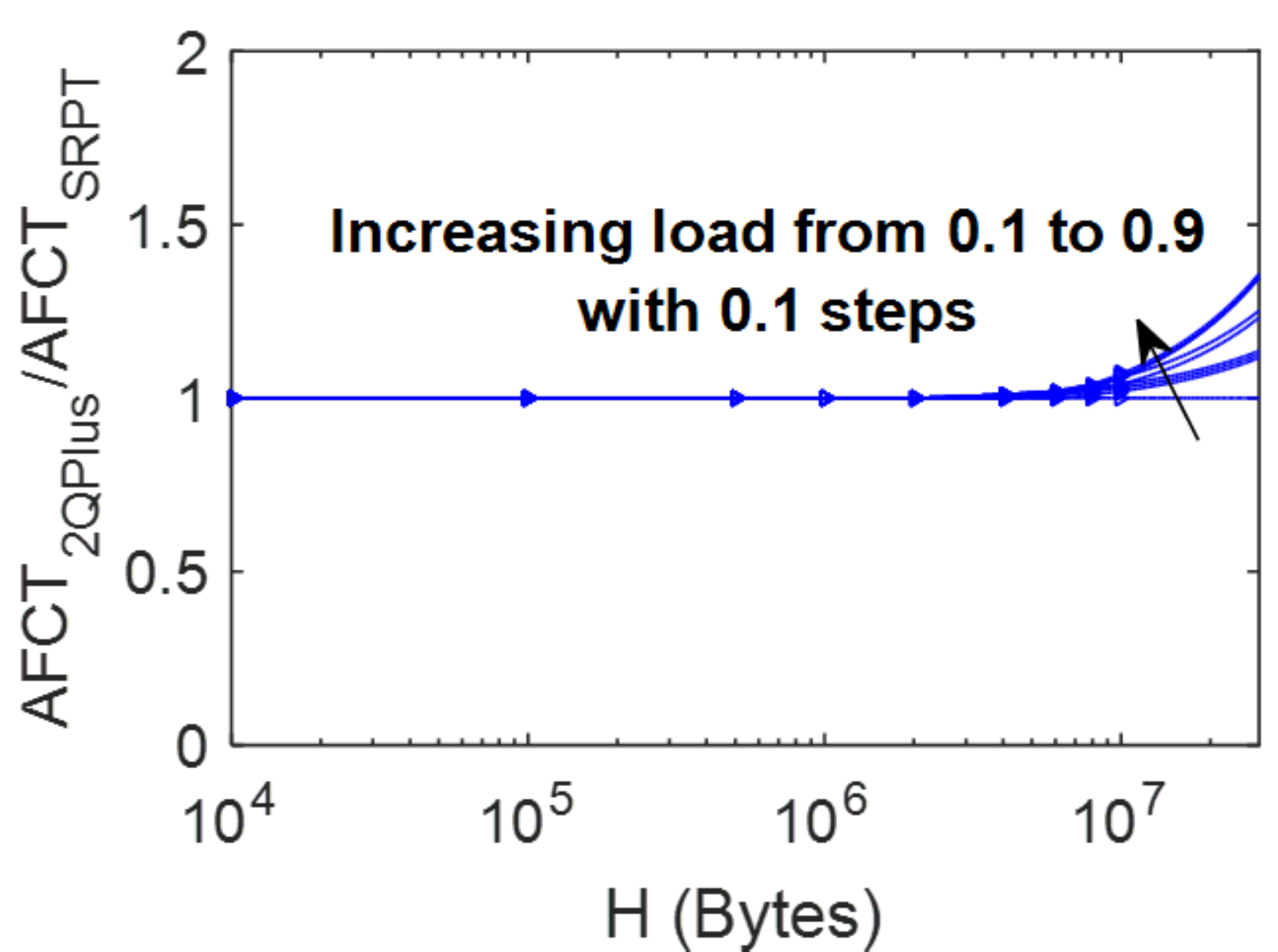}
        \subcaption{Web search’s overall flows}    
            \label{fig_afct_web_all}
    \end{minipage}
    \hfill
\begin{minipage}[b]{0.48\linewidth}
    \includegraphics[width=.9\linewidth,height=1.3in]{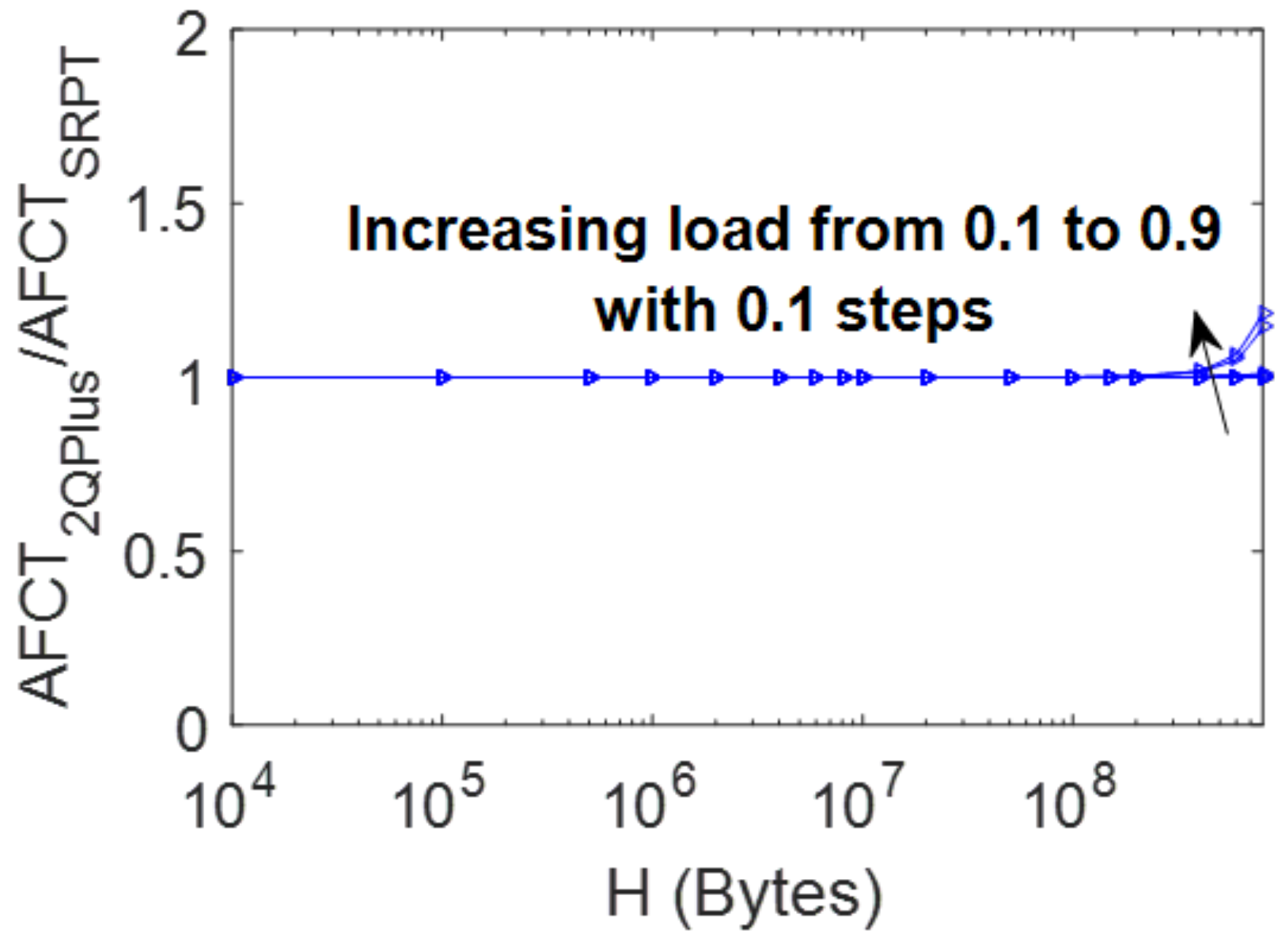}
    \subcaption{Data mining’s 2nd-class flows}    
        \label{fig_afct_data_big}
\end{minipage}
\hfill
    \begin{minipage}[b]{0.48\linewidth}
        \center
        \includegraphics[width=.9\linewidth,height=1.3in]{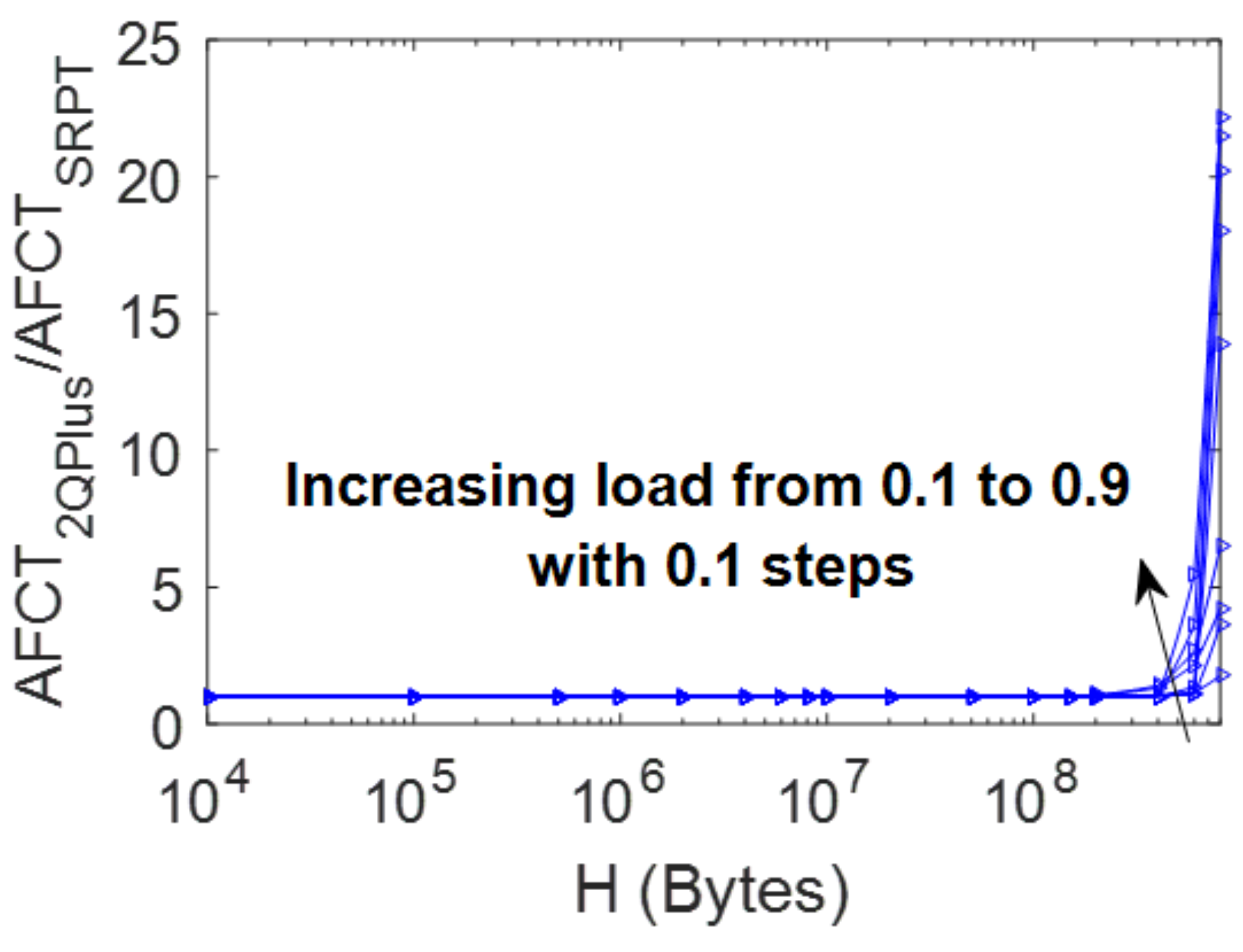}
        \subcaption{Data mining’s overall flows}    
            \label{fig_afct_data_all}
    \end{minipage}
    \caption{$AFCT_{FCFS}/AFCT_{SRPT}$ across different loads.}
    \label{fig_afct}
\end{figure}

$AFCT_{2QPlus}/AFCT_{SRPT}$ for the 2nd-class flows are shown in Fig.~\ref{fig_afct_web_big} and Fig.~\ref{fig_afct_data_big}, respectively for web search and data mining workloads. As these results illustrate, the 2nd-class flows perform roughly the same under two schemes, though there is a subtle difference between 2QPlus and SRPT especially in high loads (which causes to not have exactly equal AFCT for 2nd-class flows in both schemes). The difference is that in SRPT scheme, a 2nd-class flow that its size was originally bigger than H and currently has remaining size, $r$, $r<H$, has higher priority than any 1st-class flow with size s, where $r<s<H$, while in 2QPlus, this 2nd-class flow will always be preempted by any 1st-class flow no matter what its remaining size is. So, especially in high loads, 2nd-class flows perform slightly better in SRPT scheme. For search workload, compared to data mining workload, the number of 2nd-class flows are higher. So in high loads, overall AFCT for search workload (Fig.~\ref{fig_afct_web_all}), compared to data mining workload (Fig.~\ref{fig_afct_data_all}), will be impacted more by the 2nd-class flows. This causes to have a smaller overall difference between 2QPlus and SRPT for search workload compared to data mining workloads in high loads.

\section{Discussion}
\subsection{Impact of Routing} 
Discussions made in this paper were based on the big single-switch model of the network (which is used as the base in different prior work e.g.~\cite{pfabric,conga,qjump}). Using close-based architectures to make datacenter topologies (e.g. as in~\cite{vl2,portland,jupiter,facebook}) leads to having a non-blocking network. This non-blocking property of the network justifies the use of single-switch model for datacenter networks. Considering that, routing strategies and scheduling algorithms are usually treated as orthogonal approaches in the community. This is similar to the design of IP layer (for routing) and transport layer (for congestion control) and the impact that they can have on each other. Network designers usually design these layers independent of each other and consider them as orthogonal solutions. Although one can imagine that a cross-layer design, which considers the routing and congestion control at the same time, can potentially reach higher performance compared to having two independent deigns for these layers, separation of the design of these layers tremendously increases the modularity of the system.

Moreover, it is shown that simple routing strategies such as ECMP and random packet spraying~\cite{spray} achieve very good results for short flows in datacenter networks~\cite{conga,spray,hyline}. In other words, these approaches can distribute short flows evenly over different available links. Therefore, it can be realized that systems similar to the one described in this paper, which promotes no need for scheduling for short flows, combined with routing approaches such as packet spraying and ECMP, which are close to optimum for short flows, will provide very good performances. On the other hand, large flows can be managed independent of the approaches used for short flows both in term of scheduling and routing algorithms~\cite{hyline}. 

\subsection{From Existence of the Threshold to Calculate It}
Theorem 1 provides a subset of all values of $H$ that can guarantee the 1st-class flows with lower AFCT than in SRPT. As mentioned in section~\ref{sec_num}, maximum value of $H$ depends on three factors: 1) workload, 2) $T_{cost}$, and 3) load. In addition, for maximum value of $H$ we have: $E\left[W_{H_{max}}\right]_{FCFS}=T_{cost}$. Therefore, to choose a right value of $H$, network designers need to feed in the $T_{cost}$ value of their network, their target workload and load to the equation. Another way for choosing the desired $H$ value is to use one upper bound and one lower bound on the values of H~\cite{hyline}. In other words, in practice, reducing $H$ leads to have more flows being categorized as the 2nd class traffic and that leads to additional delays for them due to the existence of $T_{cost}$. Simply speaking, this additional delay is acceptable for flows in the 2nd class if the delay of flows in the 1st class is higher than $T_{cost}$. As it is shown in~\cite{hyline}, this will lead to a lower bound for the values of $H$. On the other hand, increasing the value of $H$, increases the number of flows in the 1st class. This leads to the increase in the delay of packets in the 1st class queue. Therefore, a practical approach is to limit the load of the 1st class traffic to a certain point. For instance, designers can fix the maximum load of the 1st class traffic to $10\%$ of the total load of the traffic. This way, there will be an upper bound on the value of the $H$. Then, $H$ can be selected from the range indicated by the lower bound and upper bound values~\cite{hyline}.

\section{Related Work}
Here, we briefly summarize some of the most relevant works to ours. Authors in ~\cite{opt} prove the optimality of SRPT over a single link, and later, ~\cite{mil} derived mean response time’s expressions of SRPT in an M/G/1 queue in the ideal case.  Using this general wisdom, ~\cite{pdq} attempts to approximate the SRPT algorithm and schedule flows in a distributed way in DCNs. In ~\cite{pdq}, end-hosts attach their flows’ information to the packets and request exact rates for sending these flows from switches. Switches gather all information and then, assign higher sending rates to higher priority flows and stop lower priority ones.~\cite{pfabric} instead, uses local-aware switches equipped with SRPT queues to schedule flows carrying their priorities locally. However, ~\cite{pase} shows that the lack of global-awareness in ~\cite{pfabric} can cause a dramatic drop in its performance. ~\cite{pase} uses a centralized controller and approximates SRPT to schedule end-hosts’ flows. To that end, end-hosts send their requests for sending flows to the controller and receive reference rates and priorities to be used for flows. ~\cite{fastpass} uses a logically centralized controller to control exact arrival times of all flows into the network. All flows send their requests to the controller asking for timeslots to come and fully use network resources.~\cite{fastpass} emulates SRPT to prioritize flows and their packets so that the higher priority ones could exploit resources first. In contrast with these papers, we have shown in this paper that based on the characteristics of the traffic in datacenter networks, there is no need to do fine scheduling for most of the flows and a simple FIFO strategy can perform very well. 

In addition to scheduling schemes focusing on datacenter networks, there are different general purpose queue management schemes and scheduling proposals. For example, GPS and its packetized version (PGPS)~\cite{gps1,gps_multi}, Stop-and-Go~\cite{sg1,sg2}, RCSP~\cite{rcsp}, HRR~\cite{hrr}, D-EDD~\cite{edd_d}, BoDe~\cite{bode}, and SharpEdge~\cite{sharpedge} try to use different techniques such as fair-queuing, TDMA-based timeframe ideas, and various traffic regulator and shapers to provide QoS guarantee over data transmissions in a network. SRPT and its other approximations focus on minimization of overall completion time of tasks, while general purpose queue management approaches concentrate more on network QoS metrics such as delay and rate guarantees. 

In addition to scheduling approaches, another branch of prior work attempt to use flow control mechanisms at the end-hosts to resolve the issue of long flow completion times. The key idea is to use algorithms in the transport layer to reduce the congestion in the network. By reducing the congestion in the network, flows can potentially achieve lower completion time and higher speeds~\cite{sprout, c2tcp, bbr, deepcc, d2tcp, xcp, c2tcp2, vegas, natcp, cubic}. Among the congestion control algorithms targeting datacenter networks, proposals such as~\cite{dctcp,d2tcp,timely,hyline} can be mentioned. TCP proposals targeting datacenter networks can alleviate the congestion in the network. Consequently, they can indirectly lead to reduction in AFCT.  

Moreover, proper routing and load balancing strategies can also improve the AFCT. A good routing scheme can distribute load evenly through available paths and reduce the congestion accordingly. Consequently, in a network with lower congestion, AFCT will decrease. Among the schemes using load-balancing and routing techniques to reduce the AFCT, ~\cite{conga,spray,presto,hedera,hyline} can be mentioned. 

As mentioned earlier, a better routing design or a better congestion control design can be seen as orthogonal solutions that can be combined with proper scheduling techniques to boost the overall performance of the system.

\section{Conclusion}
In this paper, we challenged the conventional wisdom suggesting that to minimize AFCT in DCNs, all individual flows (or most of them) should be scheduled to access network resources. We showed that it is sufficient to only categorize all flows into two coarse classes named 1st-class (consisting flows with sizes less than a certain threshold, H) and 2nd-class (including all other flows), and use a simple strict priority mechanism to serve all 1st-class flows before 2nd-class ones. Having this classification, we showed that when scheduling delays (including scheduler’s computational and communication delays) are considered, serving 1st-class flows simply in FCFS manner (i.e., without any scheduling) leads to lower AFCT of them, even when it is compared to SRPT algorithm–best-known scheduling algorithm. Interestingly, through our analysis, we showed that for different DCNs’ workload patterns and various network loads ($<1$), on average, more than 90\% of the flows can be categorized as 1st-class flows, and consequently, do not need to be scheduled to achieve low AFCT in DCNs.


\end{document}